\documentclass[aps,superscriptaddress,footinbib,twocolumn]{revtex4-2}
\usepackage[latin9]{inputenc}
\usepackage{color}
\usepackage{amsmath}
\usepackage{amssymb}
\usepackage{stmaryrd}
\usepackage{graphicx}
\usepackage{siunitx}
\usepackage{scalerel}
\usepackage{adjustbox}
\usepackage[figuresleft]{rotating}
\usepackage[unicode=true,
bookmarks=true,bookmarksnumbered=false,bookmarksopen=false,
breaklinks=false,pdfborder={0 0 1},backref=false,colorlinks=true]
{hyperref}
\hypersetup{
	linkcolor=magenta, urlcolor=blue, citecolor=blue, pdfstartview={FitH}, unicode=true}

\makeatletter

\usepackage{amsfonts}
\usepackage{tabularx}
\usepackage{dcolumn}
\usepackage{bm}
\usepackage{graphicx}
\usepackage{epstopdf}
\usepackage{times}
\hypersetup{urlcolor=blue}

\makeatother

\begin{document}
\title{Non-Abelian braiding of Fibonacci anyons with a superconducting processor}
	
\affiliation{Zhejiang Key Laboratory of Micro-nano Quantum Chips and Quantum Control, School of Physics, Zhejiang University, Hangzhou 310027, China\\
	$^2$ Center for Quantum Information, IIIS, Tsinghua University, Beijing 100084, China\\
	$^{3}$ ZJU-Hangzhou Global Scientific and Technological Innovation Center, Zhejiang University, Hangzhou 310027, China\\
	$^4$ Theoretical Physics Division, Chern Institute of Mathematics and LPMC, Nankai University, Tianjin 300071, China\\
	$^{5}$ Hefei National Laboratory, Hefei 230088, China\\
	$^{6}$ Shanghai Qi Zhi Institute, 41th Floor, AI Tower, No. 701 Yunjin Road, Xuhui District, Shanghai 200232, China}

\author{Shibo Xu$^{1}$}\thanks{These authors contributed equally to this work.}
\author{Zheng-Zhi Sun$^{2}$}\thanks{These authors contributed equally to this work.}
\author{Ke Wang$^{1}$}\thanks{These authors contributed equally to this work.}
\author{Hekang Li$^{3}$}	
\author{Zitian Zhu$^{1}$}
\author{Hang Dong$^{1}$}
\author{Jinfeng Deng$^{1}$}
\author{Xu Zhang$^{1}$}
\author{Jiachen Chen$^{1}$}
\author{Yaozu Wu$^{1}$}
\author{Chuanyu Zhang$^{1}$}
\author{Feitong Jin$^{1}$}	
\author{Xuhao Zhu$^{1}$}
\author{Yu Gao$^{1}$}
\author{Aosai Zhang$^{1}$}	
\author{Ning Wang$^{1}$}
\author{Yiren Zou$^{1}$}
\author{Ziqi Tan$^{1}$}
\author{Fanhao Shen$^{1}$}
\author{Jiarun Zhong$^{1}$}
\author{Zehang Bao$^{1}$}
\author{Weikang Li$^{2}$}
\author{Wenjie Jiang$^{2}$}
\author{Li-Wei Yu$^{4}$}
\author{Zixuan Song$^{3}$}
\author{Pengfei Zhang$^{3}$}
\author{Liang Xiang$^{3}$}
\author{Qiujiang Guo$^{3}$}
\author{Zhen Wang$^{1}$}
\author{Chao Song$^{1}$}\email{chaosong@zju.edu.cn}
\author{H. Wang$^{1, 3}$}\email{hhwang@zju.edu.cn}
\author{Dong-Ling Deng$^{2, 5}$}\email{dldeng@tsinghua.edu.cn}

\begin{abstract}       
Non-Abelian topological orders offer an intriguing path towards fault-tolerant quantum computation, where information can be encoded and manipulated in a topologically protected manner immune to arbitrary local noises and perturbations \cite{nayak2008non,stern2010non}. However, realizing non-Abelian topologically ordered states is notoriously challenging in both condensed matter and programmable quantum systems, and it was not until recently that signatures of non-Abelian statistics were observed through digital quantum simulation approaches \cite{Xu2023_digital, andersen_observation_2023, iqbal2023creation}. Despite these exciting progresses, none of them has demonstrated the appropriate  type of topological orders and associated non-Abelian anyons whose braidings alone support universal quantum computation. Here, we report the realization of non-Abelian topologically ordered states of the Fibonacci string-net model \cite{Levin2005String, Lin2021generalized} and demonstrate braidings of Fibonacci anyons featuring universal computational power \cite{Freedman2002Modular}, with a superconducting quantum processor. We exploit efficient quantum circuits to prepare the desired states and verify their nontrivial topological nature by measuring the topological entanglement entropy. In addition, we create two pairs of Fibonacci anyons and demonstrate their fusion rule and non-Abelian braiding statistics by applying unitary gates on the underlying physical qubits. Our results establish a versatile digital approach to exploring exotic non-Abelian topological states and their associated braiding statistics with current noisy intermediate-scale quantum processors.
\end{abstract}
	
\maketitle

\noindent The discovery of topological order \cite{Wen2004Quantum} has revolutionized the understanding of quantum matter based on the Landau-Ginzburg symmetry breaking paradigm \cite{Landau2013Statistical}. Different topologically ordered phases could bear exactly the same symmetries whereas showcasing topologically distinct features, such as long-range entanglement and the emergence of quasiparticles with anyonic braiding statistics \cite{Wilczek1984, Wen1990_topo, Wen1990_groundstate, Levin2005String}. They are of fundamental importance in understanding strongly correlated quantum phases of matter, and promise crucial applications in fault-tolerant quantum computing as well \cite{Kitaev2003FaultT}. Owing to their intrinsic nonlocal nature, logical code spaces immune to arbitrary local perturbations can be constructed from the topological degrees of freedom of the system, and logical operations can be implemented by creating, braiding, and fusing anyons. In general, the braiding of two anyons can be described by either Abelian or non-Abelian statistics, which leads to a complex phase factor or a unitary matrix acting on the degenerate state manifold, respectively. Non-Abelian anyons are quasiparticle excitations in topologically ordered systems that obey non-Abelian braiding statistics. They are the building blocks of topological quantum computing \cite{nayak2008non}. 

Realizing non-Abelian topologically ordered states and their associated non-Abelian anyons has been a long-sought-after goal in condensed matter physics \cite{nayak2008non,stern2010non}. Exciting progresses have been made in both theory \cite{moore1991nonabelions,Alicea2011non,ivanov2001non,Bonderson2006Detecting, clarke2013exotic, lutchyn2010majorana} and experiment \cite{Willett2013_magnetic, Banerjee2018Observation,Kasahara2018Majorana,Dolev2008Observation,Bartolomei2020Fractional,Dutta2022Distinguishing}. Yet, direct observation of non-Abelian exchange statistics has remained elusive so far. 
In recent years, significant advances have been achieved for the fabrication of programmable quantum platforms such as superconducting circuits \cite{Arute2019Quantum, Wu2021Strong, 2022Krinner_realizing, 2023Kim_evidence}, Rydberg atomic arrays \cite{Ebadi2021Quantum}, photons \cite{Zhong2020_quantum, Madsen2022_quantum}, and trapped ions \cite{Egan2021Fault, moses2023race}, giving raise to unprecedented opportunities for the synthesis and exploration of more and more complex topological quantum states \cite{Dumitrescu2022Dynamical, Zhang2022Digital, Satzinger2021Realizing, Semeghini2021_probing}. 
Along this direction, non-Abelian statistics has been observed recently by simulating the projective Ising anyons in the toric-code model \cite{andersen_observation_2023, Xu2023_digital} and creating the ground state wavefunction of non-Abelian $D_4$ topological order \cite{iqbal2023creation}. However, neither of the braidings of anyons realized in these experiments alone sustain a universal gate set. The Ising anyons is related to the Witten-$SU{\left( 2 \right)}$-Chern-Simons theory at level $k=2$, where the $SU{\left( 2 \right)}$ model is computational universal for $k=3$ or $k \ge 5$ \cite{Hormozi200704Topological}. 
For the quantum double model of the finite group (including $D_4$), the gate set realized by braiding is finite and is not universal \cite{Etingof200801Braid}. The demonstration of non-Abelian anyons suitable for universal quantum computation has evaded experiments hitherto. 

\begin{figure*}[htb]
\includegraphics[width=1.0\linewidth]{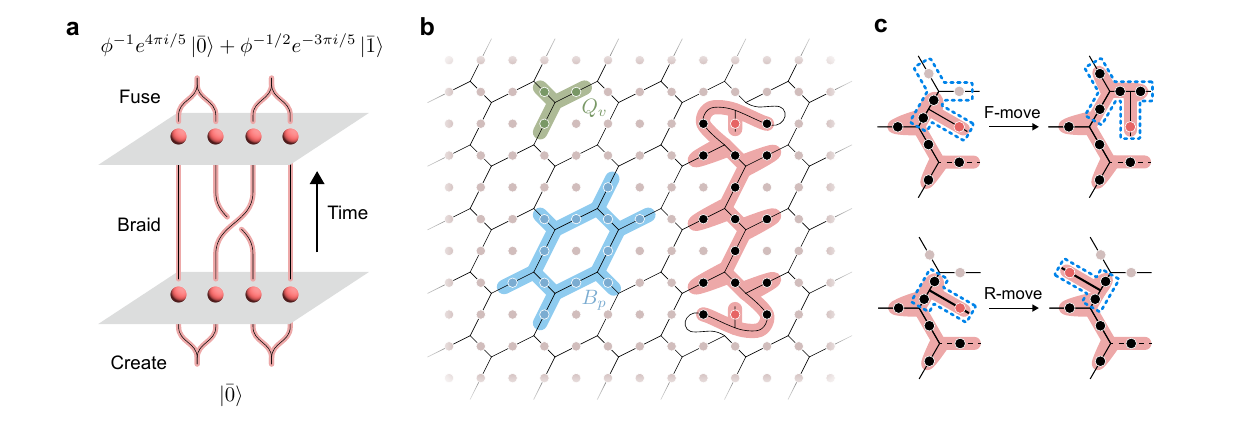}
\caption{\label{fig1} \textbf{The Fibonacci anyon  and string-net model.} \textbf{a}, World line of braiding Fibonacci anyons. We create two pairs of Fibonacci anyons from vacuum, braid the middle two, and then fuse them. In terms of topological quantum computing, such a braid will transfer the initial logical state $\left| {\bar 0} \right\rangle$ to the logical state $|\Psi\rangle_L={\phi ^{ - 1}}{e^{4\pi i/5}}\left| {\bar 0} \right\rangle  + {\phi ^{ - 1/2}}{e^{ - 3\pi i/5}}\left| {\bar 1} \right\rangle$, which can be detected by measuring the fusion results of the two pairs of anyons.  
\textbf{b}, The Fibonacci string-net model is defined on a honeycomb lattice, which in turn is constructed out of the underlying square lattice that depicts the geometry for the transmon qubits of our quantum processor. 
The $Q_v$ and $B_p$ operators are three- and twelve-body projectors acting on the qubits associated with each vertex and plaquette as highlighted in olive and blue, respectively.
A pair of Fibonacci anyons can be created at the endpoints (red dots) of an open string operator (coral line), which can be extended and turned around by F- and R-moves.
\textbf{c}, Effects of F-move (up) and R-move (down). The F-move (R-move) operator acts on five (three) qubits circled by dash lines, which extends (adjusts the direction of) the string operator and moves the Fibonacci anyon along (across) the plaquettes. See Methods and Supplementary Information I.E for details.}
\end{figure*} 

 Here, we report the experimental realization of the non-Abelian topologically ordered states of the Fibonacci string-net model \cite{Levin2005String, Lin2021generalized}, which is predicted to host Fibonacci anyons carrying universal computational power (Fig. \ref{fig1}\textbf{a}), with 27 superconducting transmon qubits. We upgrade our device through optimizing device fabrication and controlling process, and execute efficient quantum circuits obtained by variational algorithms to prepare the desired non-Abelian ground state of the string-net Hamiltonian. We measure the multi-body vertex and plaquette operators, yielding average expectation values of $0.94$ and $0.58$, respectively. The topological order of the prepared states is characterized by measuring the topological entanglement entropy, whose averaged value reaches $-0.82$, which is well below zero (for topologically trivial state) and $-0.69$ (for the $Z_2$ topologically ordered toric code state). In addition, we create two pairs of Fibonacci quasiparticle excitations by acting string operators on the prepared ground state and demonstrate their nontrivial mutual statistics by braiding them with sequences supporting universal single-qubit logic gates. We extract the characterizing monodromy matrix and quantum dimension of the Fibonacci anyon from the measured fusion results, which unambiguously indicates  that the quasiparticle excitations created in our experiment are indeed Fibonacci anyons.
	
\vspace{.5cm}

\noindent\textbf{\large{}Framework and experimental setup}
\noindent We consider the Fibonacci string-net model---the Levin-Wen model---which is the simplest string-net model supporting braiding-universal topological quantum computing \cite{Levin2005String}. The corresponding Hamiltonian is defined on a honeycomb lattice with spins living on the edges (Fig. \ref{fig1}\textbf{b}):
\begin{align}\label{eq:fixed-point-Hamiltonian}
H = -\sum\limits_v {{Q_v}}  - \sum\limits_p {{B_p}},
\end{align}
where $Q_v$ denotes the three-body vertex operator that constrains the string types meeting at a trivalent vertex, and $B_p$ denotes the twelve-body plaquette operator that measures the ``magnetic flux" through a plaquette and provides dynamics for the string-net configurations \cite{Levin2005String}. The ground state of $H$ is topologically-ordered and satisfies $\langle Q_v\rangle=\langle B_p \rangle = 1$ for all vertices $v$ and plaquettes $p$. The quasiparticle excitations are Fibonacci anyons satisfying the following fusion rule:
\begin{eqnarray}
\tau  \times \tau  = {\bf{1}} + \tau,   \label{eq:fusion-rule-Fibonacci}
\end{eqnarray}
where ${\bf{1}}$ and $\tau$ denote  the vacuum and Fibonacci anyon, respectively. They can be created and manipulated by string operators \cite{Hu2018Full}, as illustrated in Fig. \ref{fig1}{\bf{b}} and {\bf{c}}. Apparently, preparing the ground state of $H$ and manipulating Fibonacci anyons pose a significant challenge due to the intricate multi-body plaquette operators involved in the model. To overcome this difficulty, we optimize our device and exploit efficient quantum circuits, which are obtained through the variational unitary synthesis technique \cite{Nemkov2023efficient}, to prepare the desired non-Abelian ground state and use the idea of digital quantum simulation to implement creations and braidings of Fibonacci anyons (Methods and Supplementary Information III).

\begin{figure*}[htb]
\includegraphics[width=1\linewidth]{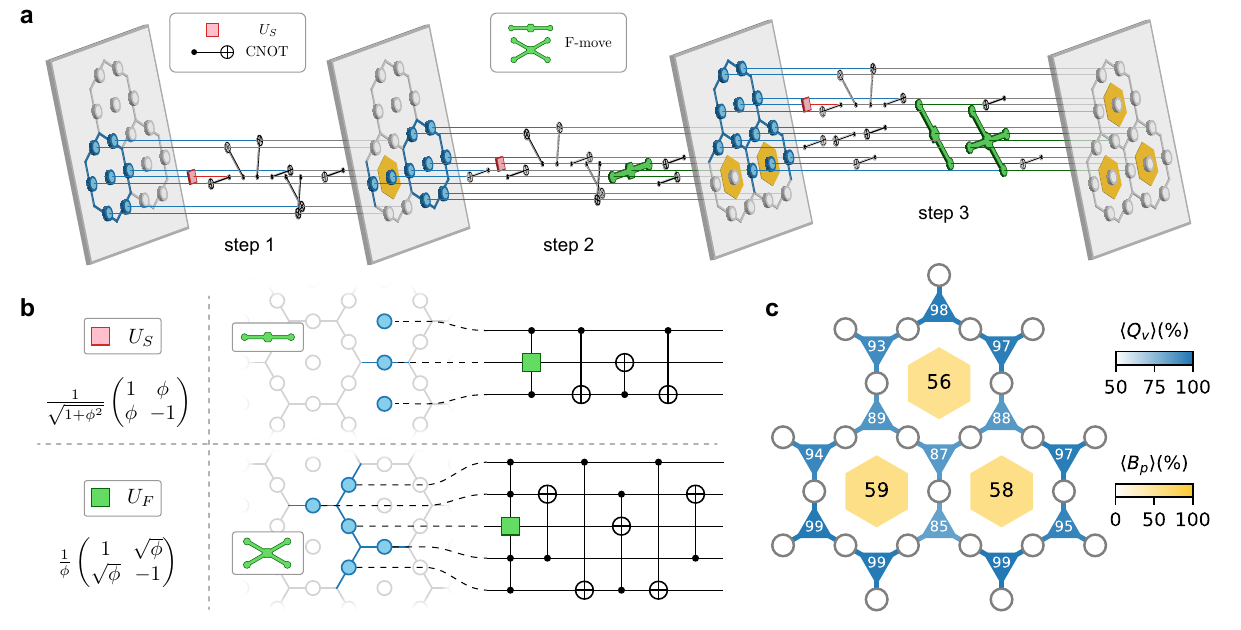}
\caption{\label{fig2} \textbf{Ground state preparation}. \textbf{a}, An illustrative quantum circuit for preparing the ground state with time flowing to the right. Starting with all qubits in $|0\rangle$ state, we project each plaquette to the ground state of the corresponding $B_p$ operator in an order indicated by the yellow hexagons. The circuit consists of single-qubit gate ${U_S}$, CNOT gates, and F-move gates, which will be decomposed further into elementary gates that are native for our quantum processor. \textbf{b}, Quantum circuits for implementing the F-moves shown in \textbf{a}. \textbf{c}, The measured $\langle Q_v\rangle$ and $\langle B_p\rangle$ for the non-Abelian topological ground state prepared in our experiment. A repetition number of 3000 (300,000) is used to obtain the probability distributions on the computational basis, which are corrected with iterative Bayesian methods \cite{DAGOSTINI1995487, Nachman2020unfolding} to mitigate readout error for obtaining $\langle Q_v\rangle$ ($\langle B_p\rangle$). 
}
\end{figure*}

Our experiments are performed on a flip-chip superconducting quantum processor with frequency-tunable transmon qubits arranged in a square lattice \cite{Xu2023_digital}.  We select $27$ neighboring qubits and construct a honeycomb lattice with three plaquettes out of the underlying square lattice (Extended Data Fig. \ref{extended_data_fig_device}). The qubits living on the edges of the honeycomb lattice are used for implementing the string-net Hamiltonian $H$, and other ones serve as ancillary qubits to facilitate the implementation of multi-qubit string operators. Arbitrary single-qubit gates can be realized for each qubit, while two-qubit controlled-Z gates can be implemented on an arbitrary neighboring qubit pair connected by a tunable coupler. Through optimizing device fabrication and controlling process, we push the median lifetime of these qubits to $117$ $\mu s$ and the median simultaneous single- and two-qubit gate fidelities around $99.96 \%$ and $99.5\%$, respectively. This enables us to successfully prepare the desired non-Abelian topological ground state of $H$ and implement the braidings of Fibonacci anyons with quantum circuits of depths up to one hundred.   See Supplementary Information III.A for the calibration procedures and detailed parameters of the device.

\vspace{.5cm}

\noindent\textbf{\large{}Ground state preparation}	

\noindent 
We prepare the ground state of $H$ by utilizing the fact that all $Q_v$ and $B_p$ are projectors commuting with each other.
Noting that the $N$-qubit product state $|0\rangle^{\otimes N}$ is an eigenstate of all $Q_v$, the ground state $|G\rangle$ can be expressed as
\begin{align}\label{eq:ground_state}
|G\rangle \propto \prod_{p}{B_p}|0\rangle^{\otimes N} = \prod_{p}{\frac{1}{{1+\phi^2}}(B_p^0+\phi B_p^1)}|0\rangle^{\otimes N},
\end{align}
where $B_p^s$ with $s\in\{0,1\}$ is a twelve-body plaquette operator and $\phi=(\sqrt{5}+1)/2$ is the golden ratio. For an independent type-0 string loop $\left| 0 \right\rangle  \otimes  \cdots  \otimes \left| 0 \right\rangle $, the $B_p^0$ operator leaves the configuration unchanged, while the $B_p^1$ operator changes it to a type-1 string loop $\left| 1 \right\rangle  \otimes  \cdots  \otimes \left| 1 \right\rangle $ according to the fusion rule ${\bf{1}} \times \tau  = \tau $ \cite{Levin2005String}. Thus, the projector $B_p$ for isolate string loops acting on the initial state $|0\rangle^{\otimes N}$ can be implemented by randomly choosing one qubit from the plaquette $p$, preparing it onto the state $\frac{1}{\sqrt{1+\phi^2}}(|0\rangle+\phi|1\rangle)$ first with a single-qubit gate $U_S = \frac{1}{\sqrt{1 + {\phi ^2}}}\left( {\begin{matrix} 1&\phi \\ \phi &{ - 1} \end{matrix}} \right)$, and then successively applying controlled-NOT (CNOT) gates on the rest qubits with the chosen qubit being the control qubit.
Furthermore, we use the F-moves to entangle different isolated loops.
The ground state can be prepared by creating isolated loops and entangling them in the honeycomb lattice layer by layer. Such an approach is efficient, in the sense that the circuit depth scales only linearly with the number of plaquettes (Methods) \cite{Liu2022Methods,Xu2023_digital}.
The quantum circuit for a step-by-step preparation of a three-plaquette ground state is sketched in Fig. \ref{fig2}\textbf{a}, which is composed of single-qubit $U_S$ gates, CNOT gates, and F-move gates.
The F-move gates can be further decomposed into multiqubit-controlled unitary gates and CNOT gates, as shown in Fig. \ref{fig2}\textbf{b}.
In our experiments, further compilations are required to fit the circuit to the nearest-neighbor geometry of our quantum device with native gates (i.e., arbitrary single-qubit gates and the two-qubit controlled-Z gate). However, direct decomposition of the five-qubit F-move is expensive and would result in a circuit with a depth of around $200$ to prepare the ground state, which is impractical to implement reliably with a system size as large as $27$ qubits for the state-of-the-art superconducting processors. We elude this dilemma by exploiting a variational approach \cite{Nemkov2023efficient} (Methods) to efficiently implement the three- and five-qubit F-move operations.
The process infidelity between the synthetic unitary $U$ and the target unitary $V$, which is defined as $1-\frac{|\text{Tr}(U^\dagger V)|}{4^n}$ with $n$ being the qubit number, is optimized to be below \num{e-5}. We note that this variational approach is device-adapted and can substantially suppress the circuit depth for implementing F-moves. Its scalability is also assured by the fact that F-moves act locally and we only need to variationally approximate F-moves up to five qubits.               
With this greatly simplified implementation of F-moves, we first prepare the ground state of $H$ step by step, as illustrated in Fig. \ref{fig2}\textbf{a}.
We measure the expectation values of $Q_v$ and $B_p$ after each step, with the results shown in the Extended Data Fig. \ref{extended_data_fig3}. While all the $Q_v$ operators are diagonal in the computational basis and hence can be directly measured in the experiment, the $B_p$ operators involve 99328 twelve-body Pauli terms in decomposition and require 290 twelve-body measurements under different Pauli bases. The average values of $Q_v$ and $B_p$ after preparing the three-plaquette ground state are $0.88$ and $0.36$, respectively.
For the preparation of the three-plaquette ground state, we can further simplify the circuit to a depth of 53 by directly targeting the final state instead of the whole unitary during the variational search, which can generate a state with an infidelity to the target state as low as \num{e-5} in theory.
We prepare the ground state with this further simplified circuit as well. The measured expectation values of $Q_v$ and $B_p$ are displayed in Fig. \ref{fig2}\textbf{c}, with average values of $0.94$ and $0.58$, respectively.
These significantly-larger-than-zero values indicate that the non-Abelian topological state prepared in our experiment indeed has a large overlap with the ideal ground state of $H$, showing the efficiency and effectiveness of our approaches. 
In the following, we use the prepared ground state through the further simplified circuit to study the exotic properties of the Fibonacci string-net model, including distinct topological entanglement entropy and braiding statistics of Fibonacci anyons. 

\begin{figure}[tb]
\includegraphics[width=1\linewidth]{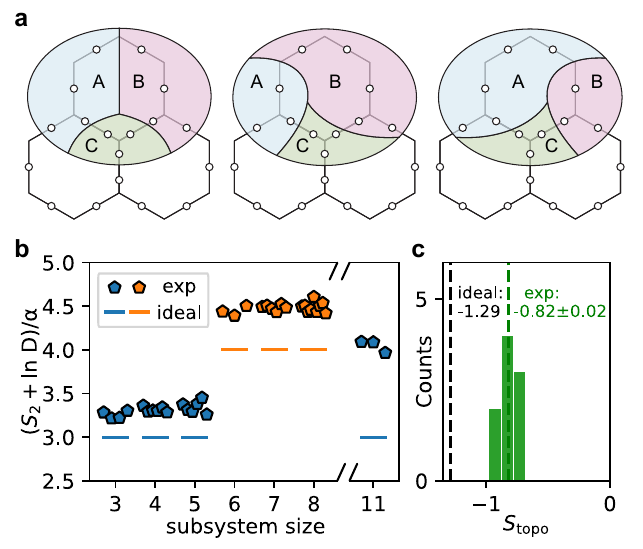}
\caption{\label{fig3} \textbf{Topological entanglement entropy}. \textbf{a}, The simply connected subregions A, B, and C used to measure topological entanglement entropy $S_{\text{topo}}$. For an eleven-qubit subsystem selected, there are three legitimate divisions  
of A, B, and C. In addition, there are three different orientations of the eleven-qubit subsystem. Thus, from a single randomized measurement on all eighteen qubits, we have nine different estimates of $S_\text{topo}$, which will converge to each other in the thermal dynamic limit.
\textbf{b}, Distribution of the rescaled second R\'enyi entropy $S_2$ measured for all the involved subsystems. Pentagon dots show experimental data and lines indicate the corresponding ideal theoretical values. 
The entropies of subsystems with boundary lengths of three and four are colored in blue and orange, respectively.
The data is rescaled to reflect the area law entanglement, where $\alpha\approx0.94$ is a constant specified by the Fibonacci string-net model \cite{Levin2006Detecting}.
\textbf{c}, Distribution of the nine extracted topological entanglement entropies, with an average value of $-0.82$ (green dashed line). 
The black dashed line indicates the ideal theoretical value.}
\end{figure} 

\begin{figure*}[htb]
\includegraphics[width=1\linewidth]{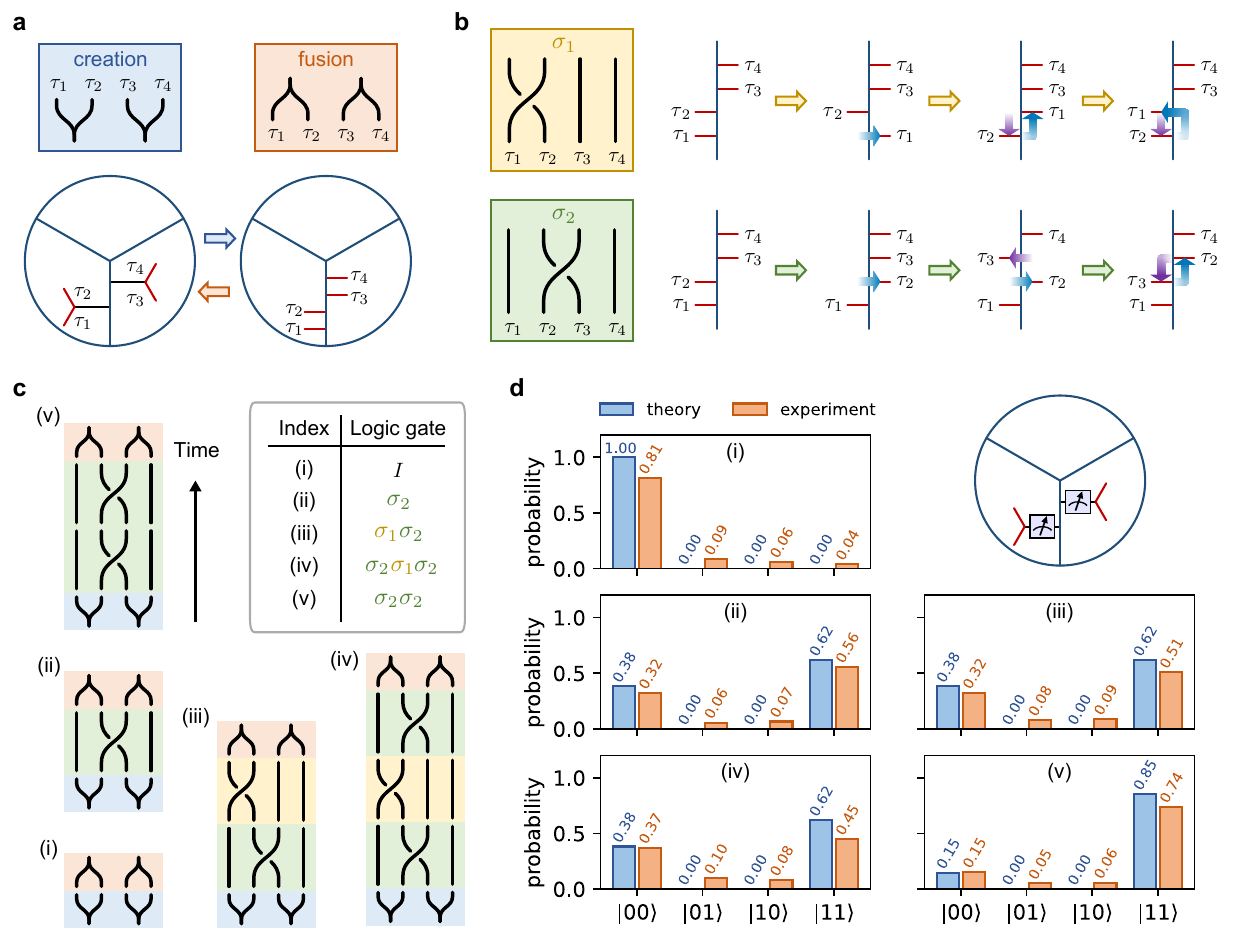}
\caption{\label{fig4} \textbf{Braiding statistics}. \textbf{a}, Creation and fusion of Fibonacci anyons, which can be described by the world lines as shown in the up panel, with time flowing from down to up. The corresponding operations in the string-net picture are shown in the lower panel, where two pairs of Fibonacci anyons can be created and fused with two F-moves and their inverses, respectively. The four anyons are labeled as $\tau_{1,2,3,4}$ and the fan sectors sketch the original hexagon plaquettes. \textbf{b}, World line representation and the corresponding string-net picture for the two braiding operations $\sigma_1$ (up) and $\sigma_2$ (down). We use R-moves to transfer the anyons across different plaquettes, and F-moves to move them along the edge (right panel).  \textbf{c}, \textbf{d}, Five braiding sequences (\textbf{c}) and the fusion results of Fibonacci anyon pairs at the end of each braiding (\textbf{d}). To demonstrate the braiding statistics, we create two pairs of Fibonacci anyons from vacuum, braid them along five different paths, and then fuse them. While the direct fusion of two anyon pairs right after their creation would lead the system back to vacuum (i), other braiding sequences will result in nontrivial fusion results (ii to v). In particular, we prepare the system into an eigenstate of $\sigma_2$ by applying $\sigma_1\sigma_2$ on the ground state (iii), which is verified by the similar fusion results observed after applying $\sigma_2\sigma_1\sigma_2$ on the ground state (iv). In addition, we can also extract the monodromy matrix by applying $\sigma_2\sigma_2$ and measuring the fusion result (v). The fusion results are obtained by measuring the two physical qubits ($Q_{(5,13)}$, $Q_{(5,9)}$) corresponding to the two string types as shown in the up right corner in \textbf{d}. See Methods and Supplementary Information I.E for details.}
\end{figure*}

\vspace{.5cm}
\noindent\textbf{\large{}Topological entanglement entropy}

\noindent To characterize the topological order of the prepared ground state $\left| G \right\rangle $, we measure its topological entanglement entropy, which is a universal constant reflecting topological properties of the entanglement that survives at arbitrarily long distances \cite{Kitaev2006Topological, Levin2006Detecting}.  
We deliberately choose three subregions $A$, $B$, and $C$ as depicted in Fig. \ref{fig3}\textbf{a}, and the topological entanglement entropy (denoted as $S_\text{topo}$) can then be obtained through 
\begin{eqnarray}\label{eq:TEE}   
S_\text{topo} =  S_A + S_B + S_C  - S_{AB}  - S_{BC}  - S_{AC}+ S_{ABC},  
\end{eqnarray}
where $AB$ indicates the union of $A$ and $B$, and $S_I$ $(I=A, B, C, AB, BC, AC, ABC)$ denotes the von Neumann entanglement entropy of a subsystem $I$: $S_I=-\text{Tr}(\rho_I\log\rho_I)$ with $\rho_I$ being the reduced density matrix. 
For the string-net model considered in our experiment, it is necessary to map the wave function to a new lattice with two qubits per boundary edge so that the partitioning can be implemented in a symmetric way \cite{Levin2006Detecting}.
From the perspective of topological quantum field theory, $S_\text{topo}$ is directly related to the total quantum dimension $D$ of the medium by $S_\text{topo}=-\log{D}$ \cite{Kitaev2006Topological, Levin2006Detecting}. For the Fibonacci string-net model, we have $D = 1 + d_\tau ^2$ with $d_\tau=\phi$ being the quantum dimension of a Fibonacci anyon. 

Directly measuring $S_\text{topo}$ requires quantum state tomography in general, which is resource-consuming and impractical for the system size considered in this work. Alternatively, one can measure the second-order R\'enyi entropy, from which $S_\text{topo}$ can be estimated up to an exponentially small deviation for the Fibonacci string-net model \cite{Flammia2009topo}. In our experiment, we adopt this approach and exploit the recently developed randomized measurement method to attain $S_\text{topo}$  \cite{Elben2018_renyi, Tiff2019probing, Satzinger2021Realizing}. 
We extend the ground state to a new lattice by copying the three qubits ($Q_{(7,11)}$, $Q_{(7,13)}$, $Q_{(5,11)}$) on the common boundary edges of the three plaquettes to the neighboring free qubits ($Q_{(9,13)}$, $Q_{(5,13)}$, $Q_{(5,9)}$) with CNOT gates, so that each common edge is associated with two qubits and can be separated into different subsystems symmetrically \cite{Levin2006Detecting} (Methods and Supplementary Information I.G). 
Our results are summarized in Fig. \ref{fig3}{\bf{b}} and \textbf{c}. In Fig. \ref{fig3}{\bf{b}}, we plot the distributions of the measured entanglement entropies of all the subsystems involved with qubit numbers ranging from three to eleven.
Ideally, the entanglement entropy of a subsystem scales linearly with its boundary, which is a reminisce of the entanglement area law \cite{Eisert2010Area} satisfied by the ground state $|G\rangle$.
In our experiment, the measured entanglement entropies are all slightly above the predicted values, which is consistent with 
numerical estimates considering 
the control and decoherence errors obtained during the calibration procedures (Supplementary Information III).
The nine extracted $S_\text{topo}$ estimates are also slightly above the predicted value as shown in Fig. \ref{fig3}{\bf{c}}. The mean value of measured $S_\text{topo}$ is $-0.82$, which is significantly lower than zero (for topologically trivial state) and $-\ln 2\approx -0.69$ (for the $Z_2$ topologically ordered state). 
This provides strong evidence for the Fibonacci topological order of the ground state $|G\rangle$.

\vspace{.5cm}

\noindent\textbf{\large{}Braiding statistics}

\noindent The topological order realized above supports a coveted type of quasiparticle---the Fibonacci anyons---whose braiding statistics can give rise to universal topological quantum computation \cite{nayak2008non}. 
To demonstrate the nontrivial braiding statistics of Fibonacci anyons, we create two pairs of them from the vacuum living on two plaquettes, as illustrated in Fig. \ref{fig4}{\bf{a}}. We then braid them following different sequences by the corresponding string operators. After braiding, we fuse them pairwise and measure the fusion outcomes to detect their braiding statistics. We encode a logical qubit into four Fibonacci anyons as $\left| {\bar 0} \right\rangle  = \left| {{{\left( {\tau  \times \tau } \right)}_{\bf{1}}},{{\left( {\tau  \times \tau } \right)}_{\bf{1}}}} \right\rangle$ and $\left| {\bar 1} \right\rangle  = \left| {{{\left( {\tau  \times \tau } \right)}_{\tau}},{{\left( {\tau  \times \tau } \right)}_{\tau}}} \right\rangle$, and denote the braiding operations of the first and middle two anyons as $\sigma_1$ and $\sigma_2$, respectively (Fig. \ref{fig4}{\bf{b}}). We note that $\sigma_1$ and $\sigma_2$ are unitary logical gates and their matrix representation can be calculated by the F- and R-moves (Methods and Supplementary Information I.B). 

Starting with the prepared ground state, we create two pairs of Fibonacci anyons labeled as $\tau_{1,2,3,4}$ by acting a short type-1 open string. In the logical space, we initialize the system into the state $\left| {\bar 0} \right\rangle$. We consider five different braiding sequences shown in Fig. \ref{fig4}{\bf{c}} and plot the corresponding measured fusion results in Fig. \ref{fig4}{\bf{d}}: (i) without braiding. As shown in Fig. \ref{fig4}{\bf{d}}(i), we measure a probability of $0.81$ and $0.04$ for both pairs fusing to $\mathbf{1}$ and $\tau$ respectively, which confirms the theoretical prediction that anyons created in pairs from vacuum will annihilate back into vacuum without braiding; (ii) braiding of the middle two Fibonacci anyons once. This will change the fusion output for both pairs, resulting in a superposition state in the logical space as  ${\sigma _2}\left| {\bar 0} \right\rangle  = {\phi ^{ - 1}}{e^{4\pi i/5}}\left| {\bar 0} \right\rangle  + {\phi ^{ - 1/2}}{e^{ - 3\pi i/5}}\left| {\bar 1} \right\rangle $. In our experiment, the measured probabilities of the two pairs after braiding fusing to $\bf{1}$ and $\tau$ are $0.32$ and $0.56$, respectively. This agrees with the theoretical prediction and verifies the non-Abelian fusion rule in Eq. (\ref{eq:fusion-rule-Fibonacci}); (iii-iv)  preparation and verification of a logical eigenstate of $\sigma_2$ through braidings. From the Yang-Baxter equation \cite{Yang1991Braid, kitaev2006anyons}, $\sigma_1\sigma_2|\bar{0}\rangle$ is an eigenstate of $\sigma_2$. This is verified by our experimental result that the difference between the fusion results before and after implementing an additional $\sigma_2$ on $\sigma_1\sigma_2|\bar{0}\rangle$ is insignificant, as manifested in Fig. \ref{fig4}{\bf{d}} (iii) and (iv); (v) braiding of the middle two Fibonacci anyons twice, which provides information about the monodromy matrix $M$ that characterizes the mutual statistics of Fibonacci anyons from the perspective of modular tensor category theory \cite{Lin2021generalized}.  The elements of $M$ can be written in the form of logical observable as ${M_{\tau \tau}} = \left\langle {\bar 0} \right|{\sigma _2}{\sigma _2}\left| {\bar 0} \right\rangle $, where ${M_{{\bf{11}}}}$, ${M_{{\bf{1}}\tau }}$, and ${M_{\tau {\bf{1}}}}$ equal $1$ since the braiding with the vacuum $\bf{1}$ does not change the fusion results. From the experimental result shown in Fig. \ref{fig4}\textbf{d}(v), we obtain that ${M_{\tau \tau}} =-0.39$, which agrees well with the theoretical value of $-1/{\phi ^2} \approx -0.38$. 
The measured quantum dimension of the Fibonacci anyon is ${\phi _{\exp }} = 1.60$, very close to the ideal value $d_\tau=\phi\approx 1.618$. This gives a piece of clear evidence that the quasiparticle excitations we created in the experiment are indeed Fibonacci anyons. 

We mention that the braidings carried out in our experiment involve no Hamiltonian dynamics of quasiparticle excitations. As a result, they are not endowed with topological protection that naturally arises from an energy gap separating the many-body degenerate ground states from the low-lying excited states. This is distinct from 
conventional protocols for braiding anyons \cite{nayak2008non} and therefore 
our experiment is more of a quantum simulation of braiding Fibonacci anyons in this sense.  
This explains as well the evident small deviations between the experimentally measured fusion results after braidings and the ideal theoretical predictions. Without topological protection, inevitable experimental imperfections including gate errors and limited  coherence time would cause a sizable infidelity for the final states after braidings. 
To leverage the Fibonacci anyons in our experiment for topologically protected quantum computing, an active error correction procedure such as the Fibonacci Turaev-Viro code \cite{Schotte202204Quantum} must be enforced during the braiding process.
We leave this interesting and important topic for future study.

\vspace{.5cm}

\noindent \textbf{\large Conclusion and outlook}

\noindent 
In summary, we have experimentally prepared the ground state of the Fibonacci string-net model with non-Abelian topological order on a programmable superconducting quantum processor.  We demonstrated the creation, braiding, and fusion of Fibonacci anyons by applying appropriate string operators on the prepared ground state. Unlike Ising-type anyons \cite{Xu2023_digital, andersen_observation_2023} and those related to the $D_4$ topological order \cite{iqbal2023creation}, the Fibonacci anyons demonstrated in our experiment support universal topological quantum computing. Combined with the potential inclusion of the error correction procedure \cite{Schotte202204Quantum} in the future, our results pave an alternative path towards fault-tolerant quantum computation. 

The controllability of the superconducting platform and the effectiveness of our variational approach in simplifying the quantum circuits demonstrated in our experiment open up several new avenues for future studies of other exotic topologically-ordered states of matter, as well as their related non-Abelian quasiparticle excitations with peculiar braiding statistics. In particular, it would be interesting and important to implement the generalized string-net models that break tetrahedral \cite{Hahn2020Generalized} or time-reversal symmetry \cite{Lin2021generalized}, admit symmetry enriched topological orders \cite{Heinrich2016Symmetry}, and others described by unitary fusion categories with fusion multiplicities \cite{Barter2022Computing}. 
Experimental realizations of such topologically ordered non-Abelian states would not only deepen our understanding of these unconventional phases of matter, but also provide valuable guidance for potential applications.

\vspace{.5cm}
\noindent\textbf{\large{}Methods}{\large\par}
\setcounter{figure}{0}
\renewcommand{\theHfigure}{A.Abb.\arabic{figure}}
\renewcommand{\figurename}{Extended Data Fig.}

\noindent\textbf{Fixed-point wavefunction}
\noindent    The Hamiltonian in Eq. (\ref{eq:fixed-point-Hamiltonian}) of the string-net model is designed to capture the most essential fixed-point wavefunctions, which are superpositions of various string-net configurations. These configurations are characterized by the geometry and the types of an ensemble of strings. The fixed-point wavefunction captures the universal properties of the string-net condensed phases in (2+1) dimensions, which can describe all so-called ``doubled'' topological phases. Here, we present the exact ground state wavefunction in the string-net picture and the corresponding quantum state simulated on physical qubits. Denoting the wavefunction as $\Phi$, it is uniquely specified by the following four local constraints \cite{Levin2005String}:
\begin{subequations}\label{eq:local-constraints}
\begin{align}\label{eq:supp-local-constraints-bending}
\Phi \left(\raisebox{-0.3cm}{\includegraphics[scale=1.0]{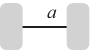}} \right) = \Phi \left( \raisebox{-0.3cm}{\includegraphics[scale=1.0]{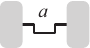}} \right),
\end{align}

\begin{align}\label{eq:local-constraints-loop}
\Phi \left( \raisebox{-0.3cm}{\includegraphics[scale=1.0]{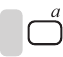}} \right) = {d_a} \Phi \left( \raisebox{-0.3cm}{\includegraphics[scale=1.0]{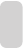}} \right),
\end{align}

\begin{align}\label{eq:local-constraints-bubble}
\Phi \left(\raisebox{-0.52cm}{\includegraphics[scale=1.0]{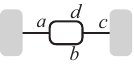}} \right) = {\delta _{ac}} \Phi \left( \raisebox{-0.52cm}{\includegraphics[scale=1.0]{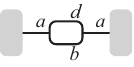}} \right),
\end{align}

\begin{align}\label{eq:local-constraints-f}
\Phi \left( \raisebox{-0.32cm}{\includegraphics[scale=1.0]{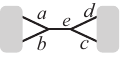}} \right) = \sum\limits_f {F_{cdf}^{abe}\Phi \left( \raisebox{-0.3cm}{\includegraphics[scale=1.0]{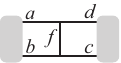}} \right)},
\end{align}
\end{subequations}
where the shaded regions represent arbitrary string-net configurations. $d_a$ is the quantum dimension of string type $a$ and the six-index tensor $F$ has a one-to-one correspondence to the doubled topological phases. Since we only consider the self-dual model in this work, all the string configurations discussed here are unoriented. The wavefunction $\Phi$ is precisely the ground state of the Hamiltonian in Eq. (\ref{eq:fixed-point-Hamiltonian}). 

According to these local constraints, the general representation of $\Phi$ can be exactly calculated for any string-net configurations. For a given geometry $g$, the wavefunction $\Phi$ becomes a function of string types $\left\{ s \right\}$. For example,
\begin{align}\label{eq:phi-one-loop}
\Phi \left( {g = \raisebox{-0.1cm}{\includegraphics[scale=1.0]{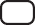}},{s_1} = a} \right) = \Phi \left( {\raisebox{-0.12cm}{\includegraphics[scale=1.0]{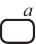}}} \right) = {d_a}\Phi \left( \text{vacuum} \right) = {d_a}.
\end{align}
where $\Phi \left( {\text{vacuum}} \right) = 1$ following the notation of Ref. \cite{Lin2021generalized}. One can also calculate the amplitudes of different string types on two independent loops:
\begin{align}\label{eq:phi-two-loop}
\Phi \left( {g = \raisebox{-0.1cm}{\includegraphics[scale=1.0]{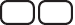}},{s_1} = a,{s_2} = b} \right) = \Phi \left( {\raisebox{-0.12cm}{\includegraphics[scale=1.0]{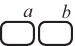}}} \right) \nonumber \\  = {d_a}{d_b}\Phi \left( {\text{vacuum}} \right) = {d_a}{d_b}.
\end{align}
From these two examples, we see that the wavefunction $\Phi$ can be recognized as a function to represent the linear relations between different string-net configurations. Once the geometry of the configuration is determined, it becomes a function of string types.

In the quantum circuit scheme, we simulate the linear relations described by the symbol $F$, which corresponds to the multi-controlled unitary gates as shown in Fig. \ref{fig2}\textbf{b}. From Eq. (\ref{eq:local-constraints-f}), the F-move changes the type of one string according to its four connected strings, which is a five-qubit gate. We can use a simplified quantum circuit to realize the F-move when there is some prior information on the string-net configuration, as shown in Fig. \ref{fig2}. Here we denote the quantum circuit corresponding to the complete and simplified F-move as $C_F$ for brevity, whereas a detailed description can be found in the Supplementary Information II.B.
Now we give the quantum state that simulates the state $\Phi$ with the geometry $g=$ one loop. For the Fibonacci string-net model, ${\Phi _1}\left( {{s_1} = 0} \right) = d_0 = 1$, and ${\Phi _1}\left( {{s_1} = 1 } \right) = {d_1 } = \phi $ according to Eq. (\ref{eq:phi-one-loop}). The corresponding normalized quantum state reads \cite{Buerschaper200902Explicit, Gu200902Tensor}
\begin{align}\label{eq:state-one-loop}
\left| G_1 \right\rangle  = {1 \over {\sqrt {1 + {\phi ^2}} }}\left( {\left| 0 \right\rangle  + \phi \left| 1 \right\rangle } \right) = {U_S}\left| 0 \right\rangle,
\end{align}
where ${U_S} = \frac{1}{\sqrt {1 + {\phi ^2}} }\left( {\begin{matrix} 1&\phi \\ \phi &{ - 1} \end{matrix}} \right)$. More precisely, the state $\Phi$ under the geometry of one isolate loop is 
\begin{align}\label{eq:phi-one-loop-map}
\Phi \left( {g = \raisebox{-0.1cm}{\includegraphics[scale=1.0]{eq/loop-noenv.pdf}}} \right) = {\Phi _1}\left( {{s_1}} \right) = {{\left\langle {{s_1}} \right.\left| {{G_1}} \right\rangle } \over {\left\langle 0 \right.\left| {{G_1}} \right\rangle }},
\end{align}
where ${\left\langle 0 \right.\left| {{G_1}} \right\rangle }$ on the denominator is for the consistence with $\Phi \left( {\text{vacuum}} \right) = 1$ \cite{Lin2021generalized}.

Similarly, we can simulate the wavefunction $\Phi$ of two independent loops with $\left| {{G_2}} \right\rangle  = {U_S}\left| 0 \right\rangle  \otimes {U_S}\left| 0 \right\rangle $. Now, we consider a more complex geometry of two connected loops:
\begin{align}\label{eq:phi-two-connect-loop}
\Phi \left( \raisebox{-0.1cm}{\includegraphics[scale=1.0]{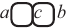}} \right) & = \sum\limits_j {F_{abj}^{bac}\Phi \left( \raisebox{-0.12cm}{\includegraphics[scale=1.0]{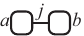}} \right)} \nonumber \\ &  = \sum\limits_j {F_{abj}^{bac}{\delta _{j0}}\Phi \left( {{\raisebox{-0.12cm}{\includegraphics[scale=1.0]{eq/loop-ab-noenv.pdf}}}} \right)} \nonumber \\ &  = {F_{ab0}^{bac}{d_a}{d_b}} .
\end{align}
The corresponding quantum state in the quantum circuit scheme is 
\begin{align}\label{eq:state-Dirac-two-connect-loop}
\left| {{G_{abc}}} \right\rangle & = {C_F}\left| {{G_3}} \right\rangle  = {C_F}\left( {\left| {{G_2}} \right\rangle  \otimes \left| 0 \right\rangle } \right) \nonumber \\ & = {C_F}\left( {{U_S}\left| 0 \right\rangle  \otimes {U_S}\left| 0 \right\rangle  \otimes \left| 0 \right\rangle } \right),
\end{align}
where $\left| {{G_2}} \right\rangle $ corresponds to the  configuration of two isolate string loops $a$ and $b$, $\left| {{G_3}} \right\rangle $ corresponds to the configuration of two string loops $a$ and $b$ connected by string $j$, and $C_F$ is the quantum circuit corresponds to $F_{ab0}^{bac}$ in Eq. (\ref{eq:phi-two-connect-loop}).

A simplified string-net wavefunction representation under the geometry shown in Fig. \ref{fig2} can be expressed as 
\begin{align}\label{eq:phi-3-palquette}
& \Phi\left(\raisebox{-0.7cm}{\includegraphics[scale=1.1]{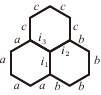}}\right)\nonumber\\
& = F_{i_1 i_2 b}^{ca i_3}\Phi\left(\raisebox{-0.75cm}{\includegraphics[scale=1.1]{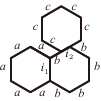}}\right) \nonumber\\
& = F_{i_1 i_2 b}^{ca i_3}F_{bc0}^{cb i_2}\Phi\left(\raisebox{-0.8cm}{\includegraphics[scale=1.1]{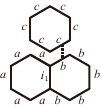}}\right) \nonumber\\
& = F_{i_1 i_2 b}^{ca i_3}F_{bc0}^{cb i_2}F_{ab0}^{ba i_1}\Phi\left(\raisebox{-0.84cm}{\includegraphics[scale=1.1]{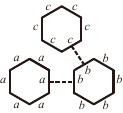}}\right) \nonumber\\
& = F_{i_1 i_2 b}^{ca i_3}F_{bc0}^{cb i_2}F_{ab0}^{ba i_1} d_a d_b d_c.
\end{align}
In the quantum circuit scheme, the initialization of an independent polygon can be realized by implementing $U_S$ first, and then entangling the rest strings of this polygon by CNOT gates controlled by the qubits to which $U_s$ applies. We call this operation as ``copy'' since this operation prepares multiple qubits to the same quantum state corresponding to the same string type. To merge the separated hexagons, we use the type-$0$ strings to connect them and use F-moves to obtain the desired geometry, similar to the process shown in Eq. (\ref{eq:phi-two-connect-loop}).
As shown in Fig. \ref{fig2}, the first F-move $F_{ab0}^{ba i_1}$ and the second F-move $F_{bc0}^{cb i_2}$ are realized by a three-qubit gate since there are two repetitive indexes for each of them. The last F-move $F_{i_1 i_2 b}^{ca i_3}$ is realized by a five-qubit gate as the general case. The number of edges changes in Eq. (\ref{eq:phi-3-palquette}), where we add and remove auxiliary qubits to the corresponding quantum circuit accordingly.
The decomposed circuits of other multi-qubit gates corresponding to the F-move and the methods to remove qubits are described in Supplementary Information II.B. 

\vspace{.5cm}

\noindent\textbf{Fixed-point Hamiltonian}

\noindent    Ref. \cite{Levin2005String} introduced an exactly solvable lattice spin Hamiltonian in the form of Eq. (\ref{eq:fixed-point-Hamiltonian}) with the fixed-pointed wavefunction $\Phi$ as the ground state. In this Hamiltonian, $Q_v$ operator is defined as:
\begin{align}\label{eq:Qv}
Q_v|ijk\rangle_v=\delta_{ijk}|ijk\rangle_v,
\end{align}
where the wavefunction $|ijk\rangle_v$ represents the types of three strings meeting at the vertex $v$, and the tensor $\delta_{ijk}$ corresponds to the fusion rules for specific anyons. For Fibonacci anyon, valid fusion rules are:
\begin{align}\label{eq:all-fusion-rules}
{\bf{1}} \times {\bf{1}} = {\bf{1}},{\bf{1}} \times \tau  = \tau  \times {\bf{1}} = \tau ,\tau  \times \tau  = {\bf{1}} + \tau,
\end{align}
which gives that ${\delta _{ijk}}=1$ if $ijk \in \{000,011,101,110,111\}$ and ${\delta _{ijk}}=0$ otherwise \cite{Schotte202204Quantum}.

Meanwhile, $B_p$ corresponds to the local constraints in Eq. (\ref{eq:local-constraints}) that uniquely specify the wavefunction capturing the properties of topologically ordered states. It is a sum of closed string operators describing particle and anti-particle pairs created from vacuum, moved along the edges of a plaquette (Fig. \ref{fig1}{\textbf{b}), and annihilated back to vacuum. In the Fibonacci string-net model,  $B_p$ is defined as,
\begin{align}\label{eq:Bp}
{B_p} = {1 \over {1 + {\phi ^2} }}(B_p^0 + \phi B_p^1),
\end{align}
where $s\in\{0,1\}$ represents the string types. $B_p^s$ changes the state on the six edges of the plaquette $p$ controlled by the state on the six outer links of $p$. The explicit algebraic form of $B_p^s$ is presented in the next subsection as the smallest closed string operator along the edge of one plaquette, which describes the process of creating a pair of type-$s$ anyons, moving around this plaquette, and fusing to vacuum.
      
\vspace{.5cm}

\noindent\textbf{String operators}
  
\noindent    The quasiparticle excitations live at the endpoints of the string operators. In Fig. \ref{fig1}\textbf{c}, we illustrate how the F- and R-moves extend the string operator and turn its direction, respectively. Here we give the explicit algebraic form to create, move, and fuse these quasiparticle excitations in the string-net picture. In the main text, we have mentioned that we use the tailed string-net picture \cite{Hu2018Full} where the tails represent the quasiparticle excitations located at the endpoints of the string operator. We also use this picture to conveniently describe the creation and fusion of these excitations. 
    
As shown in Extended Data Fig. \ref{fig-bps}{\bf{a}}, the creation of a pair of type-$s$ excitations from vacuum can be described as adding a short type-$s$ open string. Since one can erase or add the null (vacuum) strings at will \cite{kitaev2006anyons, Lin2021generalized}, the type-$s$ open string is connected to the string-net by the null string. Then we use one F-move to turn this configuration to a tailed string-net where the endpoints of the string operator are well-defined. The fusion of these excitations can be implemented by connecting the tails with F-move as shown in Extended Data Fig. \ref{fig-bps}{\bf{b}}. Under this framework, the algebraic form of the string operator is the same as that in Ref. \cite{Levin2005String}  for moving quasiparticles. The creation and fusion are defined near the endpoints of the string operator and exhibit some ambiguity, which is not important for our purposes since it does not affect the braiding statistics of the excitations \cite{Lin2021generalized}. For the closed string operators, the creation and fusion operation will introduce a constant factor related to the quantum dimension of type-$s$ string, as discussed in more detail in the next paragraph.

We consider the closed string operator $B_p^s$, which can be regarded as creating a pair of type-$s$ excitations from vacuum, winding them around in this plaquette, and then annihilating them to vacuum. As shown in Extended Data Fig. \ref{fig-bps-algebraic}, we create a pair of type-$s$ excitations at string $a$ with $F_{aaa'}^{ss0}$. Then we move the tail on the left around this plaquette with $F_{sa'b'}^{gba}$, $\cdots$, and $F_{sf'a'}^{laf}$. Finally, we annihilate these two excitations to vacuum with $F_{a's0}^{sa'a}$. According to the normalization convention \cite{Levin2005String}, $F_{aaa'}^{ss0} = {{{v_{a'}}} \over {{v_s}{v_a}}}{\delta _{saa'}}$ and $F_{a's0}^{sa'a} = {{{v_a}} \over {{v_s}{v_{a'}}}}{\delta _{saa'}}$, where $v$ represents the square root of the quantum dimension $d$. The product of these two terms is a constant factor $1/{d_s}$, which is eliminated by Eq. (\ref{eq:local-constraints-loop}) considering the fact that $B_p^s$ create a type-$s$ closed loop. The type-$s$ string operator on this plaquette can be  expressed as:
\begin{align}\label{eq:Wps-graphic}
\stretchleftright[800]{\Bigg\langle}{\adjustbox{valign=m}{\includegraphics[scale=1.2]{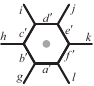}}}{\bigg |}
& {B_p^s}
\stretchleftright[800]{\bigg |}{\adjustbox{valign=m}{\includegraphics[scale=1.2]{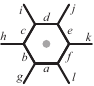}}}{\Bigg\rangle}
=  \nonumber \\ & F_{sa'b'}^{gba}F_{sb'c'}^{hcb}F_{sc'd'}^{idc}F_{sd'e'}^{jed}F_{se'f'}^{kfe}F_{sf'a'}^{laf},
\end{align}
where $B_p^s$ does not change the types of the six outgoing strings connected to the hexagon.

As the endpoints of the string operator, the tails always have a definite string type that matches the type of the corresponding string operator. Consequently, we do not attach physical qubits to the tails since there is no degree of freedom for their types. However, the tails representing the fusion results have multiple values for non-Abelian anyons and require to be captured by physical qubits as shown in Fig. \ref{fig4} and Extended Data Fig. \ref{fig-bps}. For example, in the implementation of (closed) string operator $B_p^s$, the first movement of the excitation is implemented by $F_{sa'b'}^{gba}$. Although it has six indexes, the $s$ is predetermined as the type of the simple string operator and does not occur in the quantum circuit scheme. The most complicated part in the circuit implementation of simple string operators is to apply four-qubit gates, different from the cases of preparing the ground state and the projective measurement of the plaquette operator $B_p$ where five-qubit gates are involved \cite{Liu2022Methods, Bonesteel2012Quantum}. A more detailed description to implement these multi-qubit gates is given in the Supplementary Information II.

\vspace{.5cm}

\noindent\textbf{Fusion space}

\noindent    Non-Abelian anyons have multiple fusion outputs and can be used for constructing the topologically protected logical qubits. In this work, we use four Fibonacci anyons with the vacuum total charge to encode one logical qubit. The measurement results of the logical qubit can be obtained by measuring the fusion outcomes of the first or the last two anyons. The fusion results of the first and last two anyons should be the same according to the charge conservation \cite{Xu2023_digital}. The quantum gates implemented on logical qubits are realized by the braiding operators, whose matrix representations are associated with the encoding scheme. A common calculation method is through the fusion tree notation \cite{Field2018Introduction}. 

The braiding operator $\sigma_1$ in our encoding scheme of $\left| {\bar 0} \right\rangle  = \left| {{{\left( {\tau  \times \tau } \right)}_{\bf{1}}},{{\left( {\tau  \times \tau } \right)}_{\bf{1}}}} \right\rangle$ and $\left| {\bar 1} \right\rangle  = \left| {{{\left( {\tau  \times \tau } \right)}_{\tau}},{{\left( {\tau  \times \tau } \right)}_{\tau}}} \right\rangle$ can be calculated by the R-matrix of Fibonacci theory as 
\begin{align}\label{eq:logical-s1-x}
{\sigma _1}\left| x \right\rangle  = 
\stretchleftright[800]{\bigg |}{\adjustbox{valign=m}{\includegraphics[scale=1.5]{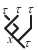}}}{\Bigg\rangle}
= 
R_{\tau \tau }^x
\stretchleftright[800]{\bigg |}{\adjustbox{valign=m}{\includegraphics[scale=1.5]{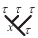}}}{\Bigg\rangle}
= R_{\tau \tau }^x\left| x \right\rangle.
\end{align}
And the braiding operator $\sigma_2$ is calculated by the F- and R-matrices of Fibonacci theory as 
\begin{align}\label{eq:logical-s2-x}
{\sigma _2}\left| x \right\rangle & = 
\stretchleftright[800]{\bigg |}{\adjustbox{valign=m}{\includegraphics[scale=1.5]{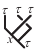}}}{\Bigg\rangle}
= \sum\limits_y F_{\tau \tau y}^{\tau \tau x}
\stretchleftright[800]{\bigg |}{\adjustbox{valign=m}{\includegraphics[scale=1.5]{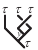}}}{\Bigg\rangle}
\nonumber \\ &
= \sum\limits_yR_{\tau \tau }^yF_{\tau \tau y}^{\tau \tau x}
\stretchleftright[800]{\bigg |}{\adjustbox{valign=m}{\includegraphics[scale=1.5]{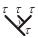}}}{\Bigg\rangle}
\nonumber \\ &
= \sum\limits_y F_{\tau \tau z}^{\tau \tau y}R_{\tau \tau }^yF_{\tau \tau y}^{\tau \tau x}
\stretchleftright[800]{\bigg |}{\adjustbox{valign=m}{\includegraphics[scale=1.5]{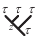}}}{\Bigg\rangle}
\nonumber \\ & 
= \sum\limits_y F_{\tau \tau z}^{\tau \tau y}R_{\tau \tau }^yF_{\tau \tau y}^{\tau \tau x}\left| z \right\rangle.
\end{align}
According to the matrix representations of the F- and R-move, one can obtain that $\sigma_1 = \left( {\begin{matrix}{e^{-4\pi i/5}} & {0}  \\ {0} & {e^{3\pi i/5}} \end{matrix} } \right)$ and ${\sigma _2} = \left( {\begin{matrix}{{\phi ^{ - 1}}{e^{4\pi i/5}}} & {{\phi ^{ - 1/2}}{e^{ - 3\pi i/5}}}  \\ {{\phi ^{ - 1/2}}{e^{ - 3\pi i/5}}} & { - {\phi ^{ - 1}}} \end{matrix} } \right)$. In the logical space, the processes (ii)-(v) shown in Fig. \ref{fig4}\textbf{c} is expressed as
\begin{subequations}\label{eq:fig4-processes}
\begin{align}\label{eq:fig4-process-2}
{\sigma _2}\left| {\bar 0} \right\rangle  = {\phi ^{ - 1}}{e^{4\pi i/5}}\left| {\bar 0} \right\rangle  + {\phi ^{ - 1/2}}{e^{ - 3\pi i/5}}\left| {\bar 1} \right\rangle ,
\end{align}

\begin{align}\label{eq:fig4-process-3}
{\sigma _1}{\sigma _2}\left| {\bar 0} \right\rangle  = {\phi ^{ - 1}}\left| {\bar 0} \right\rangle  + {\phi ^{ - 1/2}}\left| {\bar 1} \right\rangle,
\end{align}

\begin{align}\label{eq:fig4-process-4}
{\sigma _2}{\sigma _1}{\sigma _2}\left| {\bar 0} \right\rangle  = {\phi ^{ - 1}}{e^{ - 4\pi i/5}}\left| {\bar 0} \right\rangle  + {\phi ^{ - 1/2}}{e^{ - 4\pi i/5}}\left| {\bar 1} \right\rangle,
\end{align}

\begin{align}\label{eq:fig4-process-5}
\sigma _2^2\left| {\bar 0} \right\rangle  = {-\phi ^{ - 2}}\left| {\bar 0} \right\rangle  + \sqrt {3\phi  - 4} {e^{3\pi i/10}}\left| {\bar 1} \right\rangle,
\end{align}
\end{subequations}
respectively.

We notice that the Eq. (\ref{eq:fig4-process-5}) describes a three-step process defining the elements $M_{ab}$ of the monodromy matrix \cite{Lin2021generalized}: (1) create two particle-antiparticle pairs $\left( {a,\bar a,b,\bar b} \right)$ from the vacuum; (2) braid particle $a$ around particle $b$; and (3) annihilate both pairs to the vacuum, which is exactly the process shown in Fig. \ref{fig4}\textbf{c}(v). In the Fibonacci string-net model, the amplitude of the logical $\left| {\bar 0} \right\rangle $ measured in the process (v) gives the value of ${M_{\tau \tau}}$, denoted as ${M_{\tau \tau}} = \left\langle {\bar 0} \right|{\sigma _2}{\sigma _2}\left| {\bar 0} \right\rangle $. 
By the theory of modular tensor category \cite{Rowell2009Classification}, the element ${M_{\tau \tau}}$ is a real negative value taking the form of  $- 1/{d^2}$,  where $d$ is the quantum dimension of the quasiparticle excitation. In our experiment, we obtain  ${M_{\tau \tau}} \approx -0.39$, which is    
the negative square root of the experimentally measured probability for the state $\left| {00} \right\rangle $ in Fig. \ref{fig4}\textbf{d}(v). Consequently, an experimental estimation for the quantum dimension of the Fibonacci anyon is $d_\tau=\sqrt{-1/M_{\tau\tau}}\approx 1.60$.
   
\vspace{.5cm}
\noindent\textbf{Circuit implementation}

\noindent 
The original circuits for preparing the ground state and realizing different types of F-moves are compiled to fit the native gate set (i.e., arbitrary single-qubit gates and two-qubit CZ gates) and the layout geometry of the processor. 
We tackle this problem by exploiting the variational unitary synthesis technique, which can be divided into a discrete optimization part searching for the best circuit architecture and a continuous optimization part finding the best set of single-qubit rotation angles. 
In practice, we adopt the CPFlow package recently introduced in \cite{Nemkov2023efficient} to design the desired circuits.

The circuits for ground state preparation and anyon braiding in this work are composed of scalable modules and can be optimized by blocks. 
In addition, for the ground state preparation, we can further set the state vector of the ground state as the target and optimize the circuit as a whole, which further reduces the circuit depth. Before running the circuit, further alignments are executed to reduce the impact of decoherence errors, and Carr-Purcell-Meiboom-Gill (CPMG) gates are inserted to echo low-frequency noises. The experimental circuits for preparing the ground state are displayed explicitly in Supplementary Information III.B. 
\vspace{.5cm}
    
\noindent\textbf{Randomized measurement}

\noindent In our experiment, we adopt the randomized measurement (RM) method to obtain the second-order R\'enyi entropies and calculate the topological entanglement entropy \cite{Elben2018_renyi, Tiff2019probing, Satzinger2021Realizing}. This method is achieved by applying random unitaries, which are products of single-qubit unitaries sampled from the circular unitary ensemble, to the system and measuring the final states on the computational basis. 
For each instance of the random unitaries, we repeat the measurement many times to sample the probabilities of the bitstrings. The second-order R\'enyi entropy can be computed as
\begin{align}\label{eq:renyi_correlation}
\begin{split}    
S_2(\rho_A) &= -\ln(\text{Tr}(\rho_A^2)) \\
&=-\ln(2^{N_A}\sum_{w,w'}(-2)^{-H(w, w')}{\overline{P(w)P(w')}}),
\end{split}    
\end{align}
where $N_A$ and $\rho_A$ are the qubit number and density matrix of system $A$. $w, w'$ are the binary strings and $H(w, w')$ is the hamming distance between them. $P(w)$ denotes the probability of observing $w$. The average is over different instances of random unitaries in RM. 
During the calculation, we also use the iterative Bayesian unfolding scheme to mitigate measurement errors and alleviate undersampling bias (see Supplementary Information III.D). 

After preparing the ground state, we apply random unitaries to an 18-qubit system and measure its final state, from which we can obtain the second-order R\'enyi entropies of all the subsystems described in the main text. In practice, we find that an instance number of $1500$ and a sampling number of $300,000$ for each instance are required to provide a reliable estimate of the second-order R\'enyi entropy of an 11-qubit subsystem. See Supplementary Information III.C for more details on the choices of the number of instances as well as the tomography verification of the RM method with small systems.
\vspace{.5cm}

\noindent\textbf{\large{}Data availability}
The data presented in the figures and that support the other findings of this study will be publicly available upon its publication. 

\vspace{.5cm}
\noindent\textbf{Acknowledgements} We thank M. A. Levin and X.-G. Wen for enlightening discussions.
The device was fabricated at the Micro-Nano Fabrication Center of Zhejiang University.  We acknowledge the support of {Innovation Program for Quantum Science and Technology (Grant No. 2021ZD0300200 and 2021ZD0302203),} the National Natural Science Foundation of China (Grants No. 92065204, 12075128, T2225008, 12174342, 12274368, 12274367, and U20A2076), and the Zhejiang Provincial Natural Science Foundation of China (Grant No. LDQ23A040001). {H.W. is supported by the New Cornerstone Science Foundation through the XPLORER PRIZE. C.S. is supported by the Xiaomi Young Scholars Program.} Z.-Z.S., W.L., W.J., and D.-L.D. are supported in addition by Tsinghua University Dushi Program, and the Shanghai Qi Zhi Institute.

\vspace{.3cm}
\noindent\textbf{Author contributions} S.X. and K.W. ~carried out the experiments under the supervision of C.S. and H.W.. J.C. and X.Z. designed the device and H.L. fabricated the device, supervised by H.W.. Z.-Z.S.~designed the quantum circuits under the supervision of D.-L.D.. W.L., W.J., L.Y., Z.-Z.S., and D.-L.D.~conducted the theoretical analysis. All authors contributed to the experimental set-up, analysis of data, discussions of the results, and writing of the manuscript.

\vspace{.3cm}
\noindent\textbf{Competing interests}  All authors declare no competing interests.

\vspace{.3cm}
\noindent\textbf{Extended data}  

\begin{figure}[htb]
\includegraphics[width=1\linewidth]{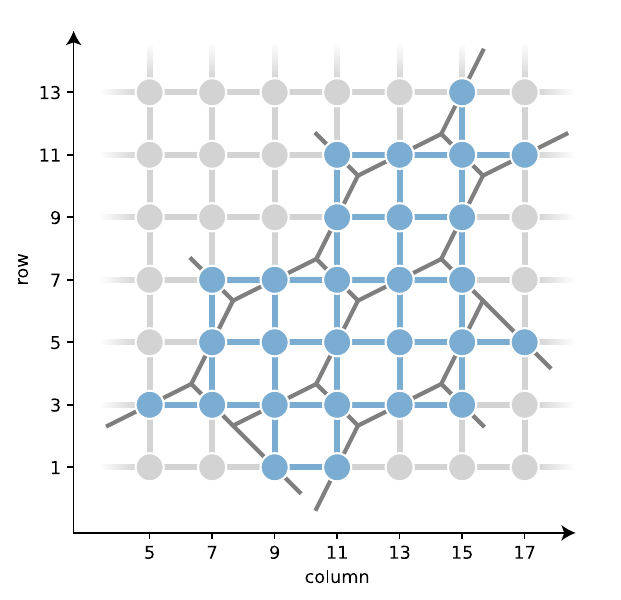}
\caption{\label{extended_data_fig_device}\textbf{The layout of the 27 qubits (blue circles) used in this experiment}, based on which we construct the honeycomb lattice. The neighboring qubits are connected with tunable couplers denoted as bars. Each physical qubit is labeled as $Q_{(i,j)}$ with $i$ $(j)$ being the row (column) index.}
\end{figure}

\begin{figure*}[htb]
\includegraphics[width=1\linewidth]{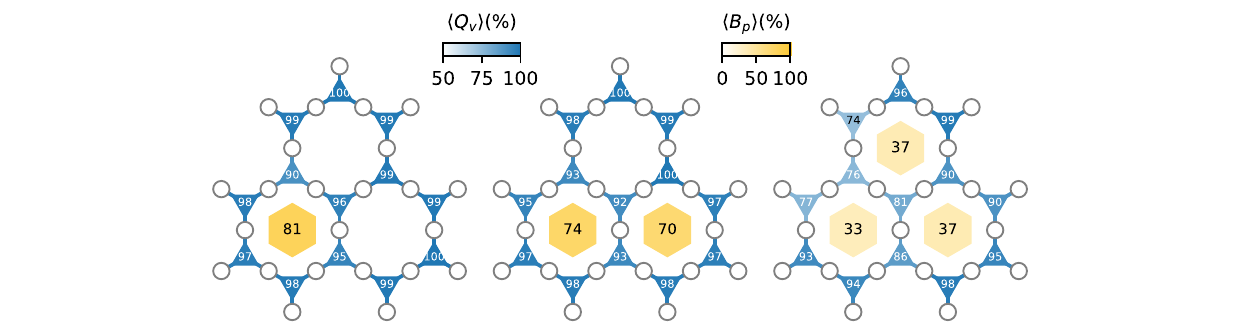}
\caption{\label{extended_data_fig3} \textbf{The measured expectation values of the vertex ($Q_v$) and plaquette ($B_p$) operators}  after step 1 (left), 2 (middle) and 3 (right) in Fig. \ref{fig2}\textbf{a}. A repetition number of 3000 (300,000) is used to obtain the probability distributions in the computational basis, which are corrected with iterative Bayesian methods \cite{DAGOSTINI1995487, Nachman2020unfolding} to mitigate readout error for calculating $\langle Q_v\rangle$ ($\langle B_p\rangle$).}
\end{figure*}

\begin{figure}[htb]
\includegraphics[width=0.8\linewidth]{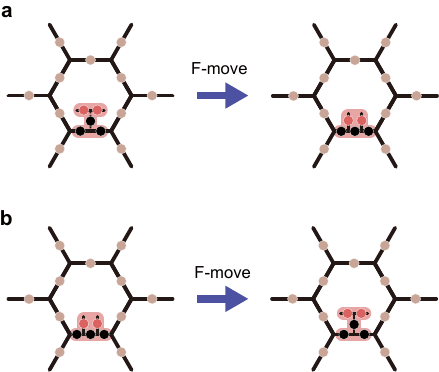}
\caption{\label{fig-bps} \textbf{Creation and fusion of the quasiparticle excitations} \textbf{a,} To create a pair of quasiparticles, we add a short string on the string-net configuration. It can be regarded as connected to the honeycomb lattice with the vacuum string, which can be arbitrarily added and removed. One F-move acting on the type-$s$ string, the type-$0$ connecting string, and the nearby edge can change the fattened lattice picture \cite{Levin2005String} to the tailed string-net picture. \textbf{b,} To annihilate (fuse) two quasiparticles, we detach the two tails with one F-move to directly connect them. The string connecting these two detached tails indicates the fusion result of these two quasiparticles.
}
\end{figure}

\begin{figure}[htb]
\includegraphics[width=1\linewidth]{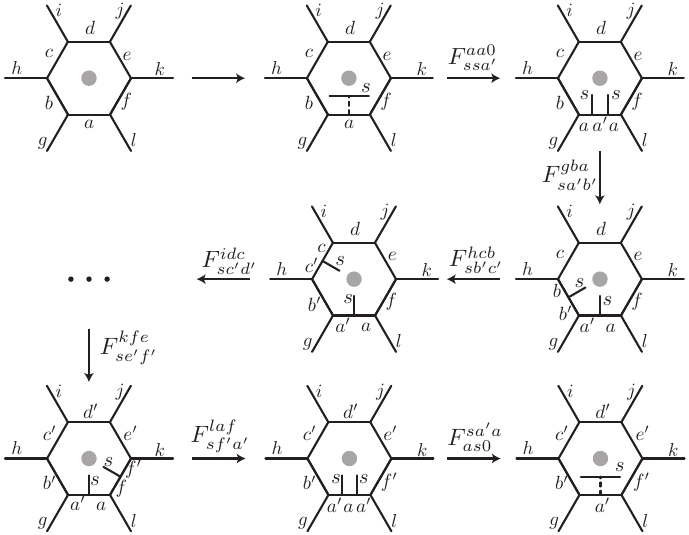}
\caption{\label{fig-bps-algebraic} \textbf{The process described by $B_p^s$ as a closed string operator.} The effect of $B_p^s$ can be understood as creating a pair of type-$s$ excitations from vacuum, moving the excitations around this plaquette, and annihilating them to vacuum. Under the tailed string-net picture, the positions of the tails clearly reveal this process. In this figure, we only move the tail initialized on the left. However, different schemes of moving these tails along this path do not change the algebraic representation of the closed string operator $B_p^s$ according to Mac Lane's coherence theorem \cite{kitaev2006anyons}.}
\end{figure}

\noindent

\clearpage
\newpage 
\onecolumngrid
\setcounter{section}{0}
\setcounter{equation}{0}
\setcounter{figure}{0}
\setcounter{table}{0}
\setcounter{page}{1}
\makeatletter
\renewcommand\thefigure{S\arabic{figure}}
\renewcommand\thetable{S\arabic{table}}
\renewcommand\theequation{S\arabic{equation}}
\renewcommand{\figurename}{FIG.}

\begin{center} 
{\large \bf Supplementary Information: {Non-Abelian braiding of Fibonacci anyons with a superconducting processor}}
\end{center} 


\section{Theory}
\label{app:theory1}

\subsection{Non-Abelian anyons}\label{sec.IA}
Anyons are quasiparticles in two dimensional space that can break the fermion-boson dichotomy \cite{Wilczek1982Quantum}. Among them, non-Abelian anyons obey non-Abelian braiding statistics: they have multiple fusion channels, and their interchanges yield unitary operations in a space spanned by topologically degenerate wavefunctions, rather than merely phase factors like fermions, bosons, and Abelian anyons \cite{nayak2008non}. Fusion rules describe the possible fusion outcomes when fusing anyons, denoted as,
\begin{align}\label{eq:fusion-rules}
a \times b = \sum\limits_c {N_{ab}^cc},
\end{align}
where $a$, $b$, and $c$ are labels of anyons, $N_{ab}^c$ is a nonnegative integer and the sum is over the complete set of labels. The quantum dimension of a specific anyon $a$ is defined as the rate at which the Hilbert space enlarges as anyons are added, denoted as $d_a$. The quantum dimension satisfies,
\begin{align}\label{eq:D-quantum-rules}
d_a d_b = \sum\limits_c {N_{ab}^cd_c}.
\end{align}

Here we briefly introduce the F- and R-moves of anyons. 
The R-move exchanges the positions of two anyons and yields a phase factor (Fig. \ref{fig-illustrate-R-move}\textbf{a}). One uncommon property of non-Abelian anyons is that the phase factor has multiple values, different from the situation of bosons, fermions, and Abelian anyons. 
According to the spin-statistics theorem, the phase factor originates from three parts.
First, the composite of $a$ and $b$ rotates clockwise by $\pi$, resulting in a phase factor of ${e^{i\pi {J_c}}}$. In the perspective of $b$, $a$ winds around $b$ anticlockwise by $\pi$, resulting in a phase factor of ${e^{-i\pi {J_a}}}$. Similarly, $b$ winds around $a$ anticlockwise by $\pi$, resulting in a phase factor of ${e^{-i\pi {J_b}}}$. Thus, the total additional phase factor of exchanging $a$ and $b$ that fusing to $c$ is $R_{ab}^c = {e^{i\pi \left( {{J_c} - {J_a} - {J_b}} \right)}}$. This phase factor is not unique since the type of $c$ is not unique in the non-Abelian case.

\begin{figure}[htb]
\includegraphics[width=0.8\linewidth]{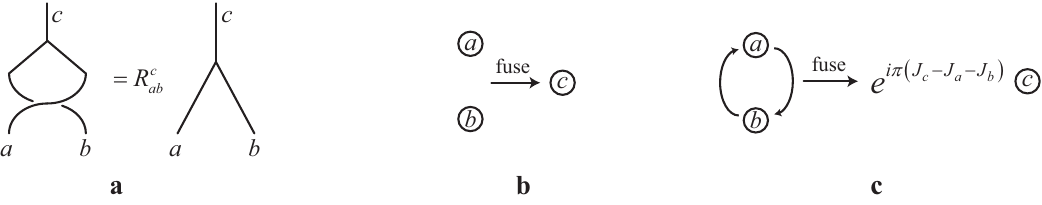}
\caption{\label{fig-illustrate-R-move} Sketch of the R-move. \textbf{a}, The R-move exchanges the positions of $a$ and $b$ before their fusion. The resulting state is related to the original state with an addition phase factor $R_{ab}^c$. \textbf{b}, The case of $a$ and $b$ fusing into $c$. \textbf{c} $a$ and $b$ exchange their positions before fusing into $c$. This process produces a quasiparticle $c$ rotating at an angle of $\pi$. According to the spin-statistics theorem, the total additional phase factor is $R_{ab}^c = {e^{i\pi \left( {{J_c} - {J_a} - {J_b}} \right)}}$ connected to multiple types of $c$.}
\end{figure}

We give a concrete description of the F-move.  Mathematically, there is an axiom satisfied by the fusion rules of an anyon model \cite{Preskill1999Lecture}:
\begin{align}\label{eq:fusion-associative}
\left( {a \times b} \right) \times c = a \times \left( {b \times c} \right),
\end{align}
which is natural since the total charge of three anyons $a$, $b$, and $c$ should be an intrinsic property of the system. Now we consider the topological Hilbert space describing the different ways to fuse $a$, $b$, and $c$ and yield a total charge $d$,
\begin{align}\label{eq:fusion-Hilbert-space}
V_{abc}^d \cong \mathop  \oplus \limits_e V_{ab}^e \otimes V_{ec}^d \cong \mathop  \oplus \limits_f V_{bc}^f \otimes V_{af}^d.
\end{align}
For example, $V_{ab}^e \otimes V_{ec}^d$ represents the Hilbert space of $a$ and $b$ fusing into $e$ first, and then $e$ and $c$ fusing into $d$. The corresponding orthonormal bases of $V_{abc}^d$ can be defined in two different ways,
\begin{align}\label{eq:two-sets-fusion-bases}
& \left| {\left( {a \times b} \right) \times c \to d;e} \right\rangle  \equiv \left| {a \times b \to e} \right\rangle  \otimes \left| {e \times c \to d} \right\rangle ,\\
& \left| {a \times \left( {b \times c} \right) \to d;f} \right\rangle  \equiv \left| {a \times f \to d} \right\rangle  \otimes \left| {b \times c \to f} \right\rangle .
\end{align}
These two sets of bases are related by a six-index tensor $F$:
\begin{align}\label{eq:F-fusion-bases}
\left| {\left( {a \times b} \right) \times c \to d;e} \right\rangle  = \sum\limits_f {F_{cdf}^{abe}\left| {a \times \left( {b \times c} \right) \to d;f} \right\rangle } .
\end{align}

\subsection{Anyonic quantum computing}

We have shown that changing fusion bases leads to a linear transformation $F$, and braiding anyons leads to a phase factor related to their total charge. Thus, we can encode the quantum information in such a topologically protected Hilbert space. For example, consider two non-Abelian anyons $a$ and $b$ with total charge $c$, the R-move produces a phase factor $R_{ab}^c$, which is basically the phase shift gate in the quantum circuit model. The topological protection of fusion space comes from the long-range entanglement between anyons far apart. 

For a condensed matter system supporting anyonic quantum computing, the low temperature can protect the quantum information encoded in the degenerate ground states from gapped excitations \cite{Kitaev2003FaultT}. In contrast, the quantum information is usually encoded in two different energy levels in the conventional quantum computer, where suppressing excitations means keeping the quantum state at the ground state $\left| 0 \right\rangle $. Meanwhile, the digital simulation of a spin model supporting anyonic excitations corresponds to an error correction scheme. Only a long string operator with a specific form can cause an undetectable error on the logical qubit \cite{nayak2008non}.

We note that in Eq. (\ref{eq:F-fusion-bases}), we can use the F-move to change the first fusion from $a \times b$  to $b \times c$. We can also change the position of anyons with the R-move when knowing their total charge. Combining these two operations allows us to change the position of anyons arbitrarily, which will provide a representation of the braid group. For example, we consider the process described by the following equation:
\begin{align}\label{eq:braid-operator}
\left| {\left( {a \times b} \right) \times c \to d;e} \right\rangle  = \sum\limits_f {F_{cdf}^{abe}\left| {a \times \left( {b \times c} \right) \to d;f} \right\rangle } & = \sum\limits_f {R_{bc}^fF_{cdf}^{abe}\left| {a \times \left( {c \times b} \right) \to d;f} \right\rangle } \nonumber \\ & = \sum\limits_f {F_{bcg}^{adf}R_{bc}^fF_{cdf}^{abe}\left| {\left( {a \times c} \right) \times b \to d;g}\right \rangle }.
\end{align}
To swap the position of anyons $b$ and $c$, we need to know the fusion result of them. Since we know that the fusion result of $a$ and $b$ is $e$, and the total charge of $a$, $b$, and $c$ is $d$, the fusion result of $b$ and $c$ is a superposition of different $f$ with the amplitude of ${F_{cdf}^{abe}}$. Then, we can use the R-move described by $R_{bc}^f$ to swap the positions of $b$ and $c$. Finally, we change the fusion basis back to first fusing $a$ and $b$ with another F-move. This operation is called the braiding operator $B$ with 
\begin{align}\label{eq:B-tensor}
B_{cdg}^{abe} = \sum\limits_f {F_{bcg}^{adf}R_{bc}^fF_{cdf}^{abe}}.
\end{align}

Denoting the exchange operation of the $i$-th and $i+1$-th particles as $\sigma_i$, it is the generator of the braid group and has the following property:
\begin{align}\label{eq:Yang-Baxter}
{\sigma _i}{\sigma _{i + 1}}{\sigma _i} = {\sigma _{i + 1}}{\sigma _i}{\sigma _{i + 1}},
\end{align}
which is known as the Yang-Baxter equation.
In our encoding scheme, we initialize the logical qubits as particle-antiparticle pairs with trivial total charge. 
The logical state is an eigenstate of $\sigma_1$ with the eigenvalue being $R_{a\bar a}^{\bf{1}}$ since 
\begin{align}\label{eq:sigma1-eigen}
{\sigma _1}{\left| {\bar 0} \right\rangle ^{ \otimes N}} = \left| {{{\left( {a \times \bar a} \right)}_{\bf{1}}},{{\left( {b \times \bar b} \right)}_{\bf{1}}}, \cdots } \right\rangle  = R_{a\bar a}^{\bf{1}}{\left| {\bar 0} \right\rangle ^{ \otimes N}}.
\end{align}
Meanwhile, the eigenstate of $\sigma_2$ is ${\sigma _1}{\sigma _2}{\left| {\bar 0} \right\rangle ^{ \otimes N}}$ with the same eigenvalue because
\begin{align}\label{eq:sigma2-eigen}
{\sigma _2}{\sigma _1}{\sigma _2}{\left| {\bar 0} \right\rangle ^{ \otimes N}} = {\sigma _1}{\sigma _2}{\sigma _1}{\left| {\bar 0} \right\rangle ^{ \otimes N}} = R_{a\bar a}^{\bf{1}}{\sigma _1}{\sigma _2}{\left| {\bar 0} \right\rangle ^{ \otimes N}},
\end{align}
which is verified by our experimental results shown in Fig. 4 in the main text.

\subsection{String-net model}

\begin{figure}[htb]
\includegraphics[width=0.7\linewidth]{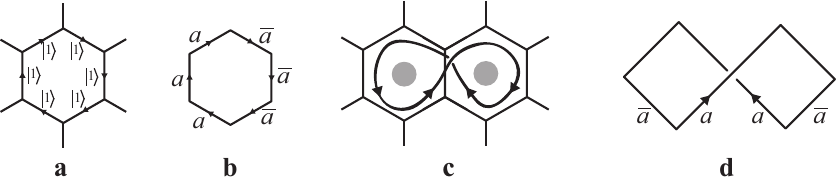}
\caption{\label{fig-fattened-lattice} String-net and the fattened lattice picture. \textbf{a}, A closed loop on the string-net. Each edge is associated with one qubit. \textbf{b}, The corresponding world line of \textbf{a}, where one particle-antiparticle pair is created and fused. \textbf{c}, A string operator in the fattened lattice picture. The strings that are not on the grid can be absorbed into the honeycomb lattice, which produces a superposition of original string-net configurations. The shaded regions in the centers of the hexagons are forbidden regions in the fattened lattice picture for the strings. \textbf{d}, The corresponding world line of \textbf{c}. Two particle-antiparticle pairs are created, one particle from the pair on the right is exchanged with the particle from the left pair in a counterclockwise sense, and then both pairs are fused. This process can also be described as braiding two anyons.}
\end{figure}

The string-net is purposed to provide a physical mechanism for ``doubled'' topological phases \cite{Levin2005String, Lin2021generalized}. In this work, we focus on the string-net model defined on the 2D honeycomb lattice, where each edge on this lattice is occupied by a spin. The degree of freedom of the spin is $N+1$, where $N$ is the number of string types. The string-net pictures can be regarded as the world lines of anyons. We use a simple example with $N=1$ to illustrate this. A close loop shown in Fig. \ref{fig-fattened-lattice}\textbf{a} is denoted by the state ${{{\left| 1 \right\rangle }^{ \otimes 6}}}$. This world line represents a process in which one particle-antiparticle pair is created and fused, as shown in Fig. \ref{fig-fattened-lattice}\textbf{b}. The direction of time is irrelevant since we  consider the original Levin-Wen model which is isotropic \cite{Levin2005String, Lin2021generalized}. 

There is also an alternative graphical representation called fattened lattice which provides a simple visual technique for understanding the Hamiltonian and quasiparticle excitations. In this picture, strings can exist on top of the original honeycomb lattice but can not cross the forbidden area in the center of each hexagon. A simple example of the closed string operator drawn in the fattened lattice picture and the corresponding world line are shown in Fig. \ref{fig-fattened-lattice}\textbf{c} and \ref{fig-fattened-lattice}\textbf{d}, respectively. The fattened lattice picture can be reduced to a linear combination of the original honeycomb lattice with some local transformations denoted as $F$ and $\Omega$, which will be introduced later.

Here we describe the basic picture of string-net condensation Ref. \cite{Levin2005String}. The topological phase occurs when the string-net is highly fluctuating and condensed. The condition of a dominating kinetic energy constraints that only string-net configurations with certain modes exist. These local constraints describe the linear relationship between amplitudes $\Phi$ of string configurations that differ by local transformations. They can be put in the following graphical form:
\begin{subequations}\label{eq:supp-local-constraints}
\begin{align}\label{eq:local-constraints-bending}
\Phi \left(\raisebox{-0.3cm}{\includegraphics[scale=1.0]{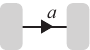}} \right) = \Phi \left( \raisebox{-0.3cm}{\includegraphics[scale=1.0]{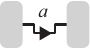}} \right),
\end{align}

\begin{align}\label{eq:supp-local-constraints-loop}
\Phi \left( \raisebox{-0.3cm}{\includegraphics[scale=1.0]{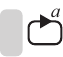}} \right) = {d_a} \Phi \left( \raisebox{-0.3cm}{\includegraphics[scale=1.0]{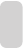}} \right),
\end{align}

\begin{align}\label{eq:supp-local-constraints-bubble}
\Phi \left(\raisebox{-0.55cm}{\includegraphics[scale=1.0]{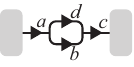}} \right) = {\delta _{ac}} \Phi \left( \raisebox{-0.55cm}{\includegraphics[scale=1.0]{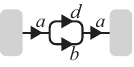}} \right),
\end{align}

\begin{align}\label{eq:supp-local-constraints-f}
\Phi \left( \raisebox{-0.3cm}{\includegraphics[scale=1.0]{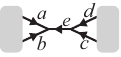}} \right) = \sum\limits_f {F_{cdf}^{abe}\Phi \left( \raisebox{-0.3cm}{\includegraphics[scale=1.0]{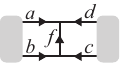}} \right)}.
\end{align}
\end{subequations}
The shaded regions represent the rest parts of the configurations, which are the same on both sides of the equations. Meanwhile, if the strings meeting at a bivalent or trivalent point do not satisfy the branching rules (corresponding to the fusion rules of the underlying anyon model), the value of the symbol $F$ related to this configuration is unphysical. We set $F=0$ in this case, which implies that the ground state configurations are required to meet the branching rules.
In this work, we only consider the self-dual model where all strings are unoriented, namely $a^* = a$. Thus, we sometimes do not draw the arrows on the string-net configurations and do not distinguish the conjugation of $F$ symbols ($F$ is real-valued). 

We note that the values of $F$ and $d$ can not be chosen arbitrarily, otherwise the local constraints of Eq. (\ref{eq:supp-local-constraints}) may be not self-consistent. Only $F$ and $d$ meeting the following conditions support the well-defined string-net condensed wavefunction (up to a trivial rescaling):
\begin{subequations}\label{eq:F-constraints}
\begin{align}\label{eq:strnet-physicality}
\text{physicality}&: \quad	F^{ijm}_{kln} \delta_{ijm} \delta_{klm^*} = F^{ijm}_{kln} \delta_{iln} \delta_{jkn^*}  
\end{align}
\begin{align}\label{eq:strnet-pentagon}
\text{pentagon equation}&:	\sum_{n=1}^N F^{mlq}_{kpn} F^{jip^*}_{mns} F^{jsn}_{lkr} = F^{jip^*}_{q^*kr} F^{r^*iq^*}_{mls} 
\end{align}
\begin{align}\label{eq:strnet-unitarity}
\text{unitarity}&:	\quad	\left(F^{ijm}_{kln}\right)^* = F^{lin}_{jkm^*} 
\end{align}
\begin{align} \label{eq:strnet-tetrahedral}
\text{tetrahedral symmetry}&: F^{ijm}_{kln} = F^{jim}_{lkn^*} = F^{lkm^*}_{jin} = F^{imj}_{k^*nl}\frac{v_m v_n}{v_j v_l}
\end{align}
\begin{align}\label{eq:strnet-normalization}
\text{normalization}&:	\quad	 F^{ii^*\bf{1}}_{j^*jk} = \frac{v_k}{v_i v_j}\, \delta_{ijk} 
\end{align}
\end{subequations}
where $ v_i = \sqrt{d_i}$. We note that the physicality [Eq. (\ref{eq:strnet-physicality})] and unitarity [Eq. (\ref{eq:strnet-unitarity})] do not apply to the general case \cite{Schotte202204Quantum}. They are purposed to avoid the nonphysical solutions and can realize topological phases that do not break time-reversal symmetry \cite{Lin2021generalized, Schotte202204Quantum}.

\subsection{Fixed-point Hamiltonian}\label{Fixed-point Hamiltonian}

The linear relationship described in Eq. (\ref{eq:supp-local-constraints}) can uniquely determine the ground state wavefunction $\Phi$, which describes the amplitude of string-net configurations. The corresponding Hamiltonian is composed of two terms:
\begin{align}\label{eq:supp-fixed-point-Hamiltonian}
H = \sum\limits_v {{Q_v}}  + \sum\limits_p {{B_p}},
\end{align}
where ${{Q_v}}$ corresponds to the fusion rules of anyons on the vertices and ${B_p} = \sum\limits_s {{a_s}B_p^s} $ on the plaquettes provides dynamics. The $s = 0,1, \cdots N$ represents the different string types. Each $Q_v$ involves three strings connected to one vertex and each $B_p$ involves six strings of a plaquette and six strings connected to this plaquette. The $Q_v$ constrains that all the strings meeting on the same vertex to satisfy the fusion rules, which can be represented as,
\begin{align}\label{eq:Qv-graphic}
{Q_v}\left| {\raisebox{-0.3cm}{\includegraphics[scale=1.0]{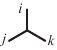}}} \right\rangle  = {\delta _{ijk}}\left| {\raisebox{-0.3cm}{\includegraphics[scale=1.0]{eq/vertex-Qv.pdf}}} \right\rangle \text{, where } {\delta _{ijk}} = 
\begin{cases}
1, & \text{if } \left\{ {i,j,k} \right\} \text{ is allowed},\\
0,    & \text{otherwise}.
\end{cases}
\end{align}
The $B_p^s$ can be easily understood in the fattened lattice picture. It adds a loop of string type $s$ cycling the shaded region, which is 
\begin{align}\label{eq:Bps-graphic}
B_p^s
\stretchleftright[800]{\bigg|}{\adjustbox{valign=m}{\includegraphics[scale=1.2]{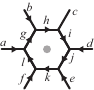}}}{\Bigg\rangle}
= 
\stretchleftright[800]{\bigg|}{\adjustbox{valign=m}{\includegraphics[scale=1.2]{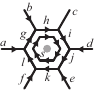}}}{\Bigg\rangle}
= \sum\limits_{mnopqr} {F_{s{m^*}r}^{a{{\lg }^*}}F_{s{n^*}m}^{bg{h^*}}F_{s{o^*}n}^{ch{i^*}}F_{s{p^*}o}^{di{j^*}}F_{s{q^*}p}^{ej{k^*}}F_{s{r^*}q}^{fk{l^*}}
\stretchleftright[800]{\bigg|}{\adjustbox{valign=m}{\includegraphics[scale=1.2]{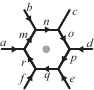}}}{\Bigg\rangle}
}.
\end{align}
The coefficients $a_s$ is selected to be ${a_s} = {d_s}/\sum\nolimits_{i = 0}^N {d_i^2} $. This choice corresponds to a topological phase with a smooth continuum limit \cite{Levin2005String}, and makes $B_p$ a projector. This can be proved by:
\begin{align}\label{eq:proof-bp-projector}
{\left( {{B_p}} \right)^2} = {\left( {\sum\limits_s {{a_s}B_p^s} } \right)^2} = {1 \over {{D^2}}}\sum\limits_{sr} {{d_s}{d_r}B_p^sB_p^r}  = {1 \over {{D^2}}}\sum\limits_{srt} {{d_s}{d_r}{\delta _{rst}}B_p^t}  = {1 \over {{D^2}}}\sum\limits_{st} {d_s^2{d_t}B_p^t}  = {1 \over D}\sum\limits_t {{d_t}B_p^t}  = \sum\limits_s {{a_s}B_p^s}  = {B_p},
\end{align}
where $D = \sum\limits_s {d_s^2} $ is the total quantum dimension of the model. This proof utilizes $B_p^sB_p^r = \sum\limits_t {{\delta _{rst}}B_p^t} $ which is proved in Ref. \cite{Levin2005String}, and ${d_a}{d_b} = \sum\limits_c {{\delta _{abc}}{d_c}} $, which is Eq. (\ref{eq:D-quantum-rules}) in multiplicity-free case. 

For the Fibonacci string-net model, the nontrivial fusion rule reads
\begin{align}\label{eq:supp-fusion-rule-Fibonacci}
\tau  \times \tau  = {\bf{1}} + \tau,
\end{align}
where ${\bf{1}}$ denotes a null string (vacuum).
We record the state of an isolate string loop as $\left| {\bf{1}} \right\rangle  + x\left| \tau  \right\rangle $. This state does not change when adding a null loop (implementing $B^0$), which is trivial. We also require that the state is unchanged (up to a rescaling factor) when adding a loop of $\tau$. The resulting state is $\left| {{\bf{1}} \times \tau } \right\rangle  + x\left| {\tau  \times \tau } \right\rangle  = x\left| {\bf{1}} \right\rangle  + \left( {x + 1} \right)\left| \tau  \right\rangle $. One solution of $x$ is ${{1 + \sqrt 5 } \over 2}$, which is the quantum dimension of the Fibonacci anyon $d_{\tau}$. The other solution is ${1 - \sqrt 5 } \over 2$, which is associated with the Yang-Lee singularity \cite{Wang2010Topological} and does not correspond to a physically realizable topological phase for string-net model \cite{Levin2005String}. 

Another important property of $B_p$ is that it ''commutes'' with the F-moves \cite{Koenig2009Exact}. In the main text, we have shown that we can change the geometry of the string-net with F-moves. 
Such deformation does not change the measurement result of $B_p$, which always corresponds to adding a type-$s$ string loop in the fattened lattice picture.
The consistency of this interpretation is guaranteed by Mac Lane's coherence theorem as long as the deformation obeys the local rules.
As an example, here we give a detailed proof for the deformation from a hexagon to a pentagon with the explicit expression given by Eq. (\ref{eq:Bps-graphic}), combined with the constraints of the F-matrix of Eq. (\ref{eq:F-constraints}). We denote the operator of adding a virtual loop of type-$s$ to a pentagon in the fattened lattice picture as $D_p^s$. The expected value of $B_p^s$ under arbitrary state $\left| \Psi  \right\rangle $ can be calculated according to Eq. (\ref{eq:Bps-graphic}), which is 
\begin{align}\label{eq:Bps-observable}
\left\langle \Psi  \right|B_p^s\left| \Psi  \right\rangle  = \sum\limits_{a \sim r} {F_{s{g^*}l}^{a{{lg }^*}}F_{s{h^*}g}^{bg{h^*}}F_{s{i^*}h}^{ch{i^*}}F_{s{j^*}i}^{di{j^*}}F_{s{k^*}j}^{ej{k^*}}F_{s{l^*}k}^{fk{l^*}}Tr\left(
\stretchleftright[800]{\Bigg\langle}{\adjustbox{valign=m}{\includegraphics[scale=1.2]{eq/phi-Bps-before.pdf}}}{\bigg |} \Psi \rangle  \langle \Psi  \stretchleftright[800]{\bigg |}{\adjustbox{valign=m}{\includegraphics[scale=1.2]{eq/phi-Bps-after.pdf}}}{\Bigg\rangle}
\right)},
\end{align}
where ``Tr'' means partial tracing over other strings (qubits) and $\sum\limits_{a \sim r} $ represent the sum over all indices from ``a'' to ``r''. Meanwhile, the matrix elements of $D_p^s$ can be calculated by:
\begin{align}\label{eq:Dps-graphic}
D_p^s
\stretchleftright[800]{\bigg |}{\adjustbox{valign=m}{\includegraphics[scale=1.2]{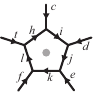}}}{\Bigg\rangle}
= 
\stretchleftright[800]{\bigg |}{\adjustbox{valign=m}{\includegraphics[scale=1.2]{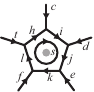}}}{\Bigg\rangle}
= 
\sum\limits_{nopqr} {F_{s{n^*}r}^{tl{h^*}}F_{s{o^*}n}^{ch{i^*}}F_{s{p^*}o}^{di{j^*}}F_{s{q^*}p}^{ej{k^*}}F_{s{r^*}q}^{fk{l^*}}
\stretchleftright[800]{\bigg |}{\adjustbox{valign=m}{\includegraphics[scale=1.2]{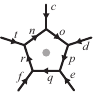}}}{\Bigg\rangle}}.
\end{align}
Thus, the expected value of $D_p^s$ on the deformed lattice corresponding to the state $F\left| \Psi  \right\rangle $ can be written as
\begin{align}\label{eq:Dps-observable}
\left\langle \Psi  \right|{F^\dag }D_p^sF\left| \Psi  \right\rangle & = Tr\left( {\sum\limits_{a \sim l} {{F^\dag }D_p^sF
\stretchleftright[800]{\bigg |}{\adjustbox{valign=m}{\includegraphics[scale=1.2]{eq/phi-Bps-before.pdf}}}{\Bigg\rangle}
\stretchleftright[800]{\Bigg\langle}{\adjustbox{valign=m}{\includegraphics[scale=1.2]{eq/phi-Bps-before.pdf}}}{\bigg |} \Psi \rangle \left\langle \Psi \right |}}
\right) \nonumber \\
& = \sum\limits_{a \sim l,t} {Tr\left( {F_{{h^*}b{t^*}}^{a{{\lg }^*}}{F^\dag }D_p^s
\stretchleftright[800]{\bigg|}{\adjustbox{valign=m}{\includegraphics[scale=1.2]{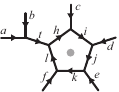}}}{\Bigg\rangle}
{\stretchleftright[800]{\Bigg\langle}{\adjustbox{valign=m}{\includegraphics[scale=1.2]{eq/phi-Bps-before.pdf}}}{\bigg |} \Psi \rangle \left\langle \Psi \right |}}
\right)} \nonumber \\
& = \sum\limits_{a \sim l,n \sim r,t} {Tr\left( {F_{s{n^*}r}^{tl{h^*}}F_{s{o^*}n}^{ch{i^*}}F_{s{p^*}o}^{di{j^*}}F_{s{q^*}p}^{ej{k^*}}F_{s{r^*}q}^{fk{l^*}}F_{{h^*}b{t^*}}^{a{{\lg }^*}}{F^\dag }
\stretchleftright[800]{\bigg|}{\adjustbox{valign=m}{\includegraphics[scale=1.2]{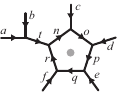}}}{\Bigg\rangle}
{\stretchleftright[800]{\Bigg\langle}{\adjustbox{valign=m}{\includegraphics[scale=1.2]{eq/phi-Bps-before.pdf}}}{\bigg |} \Psi \rangle \left\langle \Psi \right |}}
\right)} \nonumber \\
&  = \sum\limits_{a \sim r,t} {F_{s{n^*}r}^{tl{h^*}}F_{s{o^*}n}^{ch{i^*}}F_{s{p^*}o}^{di{j^*}}F_{s{q^*}p}^{ej{k^*}}F_{s{r^*}q}^{fk{l^*}}F_{{h^*}b{t^*}}^{a{{\lg }^*}}{{\left( {F_{{n^*}b{t^*}}^{ar{m^*}}} \right)}^*}Tr\left( 
\stretchleftright[800]{\Bigg\langle}{\adjustbox{valign=m}{\includegraphics[scale=1.2]{eq/phi-Bps-before.pdf}}}{\bigg |} \Psi \rangle \langle \Psi  \stretchleftright[800]{\bigg |}{\adjustbox{valign=m}{\includegraphics[scale=1.2]{eq/phi-Bps-after.pdf}}}{\Bigg\rangle}
\right)}.
\end{align}
Based on the unitarity of the F-matrix from Eq. (\ref{eq:strnet-unitarity}), ${\left( {F_{{n^*}b{t^*}}^{ar{m^*}}} \right)^*} = F_{r{n^*}m}^{ba{t^*}}$. We can also derive that $F_{{h^*}b{t^*}}^{a{{\lg }^*}} = F_{b{h^*}t}^{la{g^*}}$, $F_{s{n^*}r}^{tl{h^*}} = F_{ltr}^{{n^*}sh}$, and $F_{r{n^*}m}^{ba{t^*}} = F_{abm}^{{n^*}rt}$ according to the tetrahedral symmetry [Eq. (\ref{eq:strnet-tetrahedral})] of the Levin-Wen model. With the pentagon equation, we obtain
\begin{align}\label{eq:pentagon-D}
\sum\limits_t {F_{b{h^*}t}^{la{g^*}}F_{ltr}^{{n^*}sh}F_{abm}^{{n^*}rt}}  = F_{gbm}^{{n^*}sh}F_{lar}^{{m^*}sg} = F_{s{m^*}r}^{a{{\lg }^*}}F_{s{n^*}m}^{bg{h^*}},
\end{align}
where the second equal sign uses the tetrahedral symmetry again. Thus, the expected value of $B_p$ on the original hexagon equals the expected value of ${D_p}$ on the pentagon, which can be further reduced to the expected value of the operator that adds a virtual loop on a tadpole \cite{Bonesteel2012Quantum, Koenig2009Exact}.

\subsection{Quasiparticle excitations}

The quasiparticle excitations can be created in pairs and have a stringlike structure. The position of such a string operator is unobservable in the string-net condensed state, and its endpoints behave like independent particles. If the string operator forms a loop with no endpoints, it should commute with the Hamiltonian and become unobservable. The string loops can be directly drawn in the fattened lattice picture. Meanwhile, we need a new correspondence relationship to map the cross in fattened lattice to a linear combination of string-net configurations \cite{Levin2005String}:
\begin{subequations}\label{eq:Omega-relation}
\begin{align}\label{eq:local-constraints-wilson-loop}
\left| \raisebox{-0.3cm}{\includegraphics[scale=1.0]{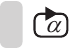}} \right\rangle = \sum\limits_i {{n_{\alpha ,i}} \left| \raisebox{-0.3cm}{\includegraphics[scale=1.0]{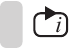}} \right\rangle},
\end{align}
\begin{align}\label{eq:local-constraints-wilson-clockwise}
\left| \raisebox{-0.4cm}{\includegraphics[scale=1.0]{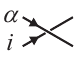}} \right\rangle = \sum\limits_{jkm} {\Omega _{\alpha ,jki}^m \left| \raisebox{-0.4cm}{\includegraphics[scale=1.0]{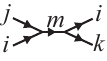}} \right\rangle},
\end{align}
\begin{align}\label{eq:local-constraints-wilson-anticlockwise}
\left| \raisebox{-0.4cm}{\includegraphics[scale=1.0]{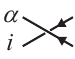}} \right\rangle = \sum\limits_{jkm} {\bar \Omega _{\alpha ,jki}^m \left| \raisebox{-0.4cm}{\includegraphics[scale=1.0]{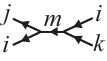}} \right\rangle}.
\end{align}
\end{subequations}
In this work, we only focus on the ``simple'' string operators where $j = k = \alpha $. More specifically, we only handle a subcategory $\left\{ {1,\tau } \right\}$ of the Fibonacci string-net model and ignore its counterpart with the opposite chirality. The Fibonacci anyon is characterized by type-$1$ simple string operators.

\begin{figure}[htb]
\includegraphics[width=0.6\linewidth]{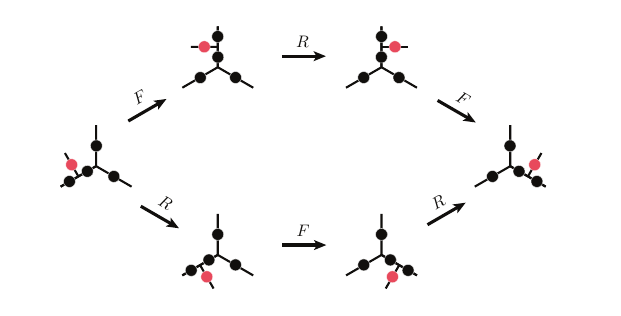}
\caption{\label{fig-hexagon-equation} Two equivalent paths to move one tail to the destination, which describes the hexagon equation for anyon models.}
\end{figure}

We have shown that the F-move commutes with the $B_p$ operator in Eq. (\ref{eq:Dps-observable}) based on the pentagon equation Eq. (\ref{eq:strnet-pentagon}). Thus, the string operator can freely move in the plaquette while remaining unobservable. A similar statement applies to crossing different plaquettes with the R-move, which is based on the hexagon equation. This relationship can be represented by the equivalence of the two processes shown in Fig. \ref{fig-hexagon-equation}, abbreviated as $RFR=FRF$. The corresponding graphical representation with the $\Omega$ notation is:

\begin{align}\label{eq:Omega-hexagon}
\stretchleftright[800]{\bigg |}{\adjustbox{valign=m}{\includegraphics[scale=0.3]{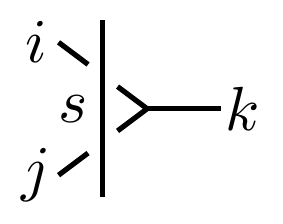}}}{\Bigg\rangle}
= 
\stretchleftright[800]{\bigg |}{\adjustbox{valign=m}{\includegraphics[scale=0.3]{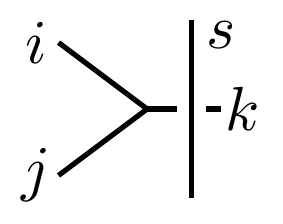}}}{\Bigg\rangle}
\end{align}

There is some ambiguity when defining the endpoints of the string operator. However, this ambiguity does not influence the mutual statistics of the quasiparticle excitations since it only depends on the topology of how string operators braid instead of the local effect at the endpoints \cite{Lin2021generalized}. In this work, we define the endpoints as the tails on the string-net configuration \cite{Hu2018Full}, which is directly reduced from an open string in the fattened lattice picture. The reduction can be performed by the linear relationship in Eq. (\ref{eq:supp-local-constraints}) and (\ref{eq:Omega-relation}). Under the tailed string-net picture, the algebraic form of the closed simple string operator is the same as that in the Ref \cite{Levin2005String}. Since the quasiparticle excitations are clearly located at the tails, we can conveniently express the mutual statistics with this figure.

Here we give an example of the shortest ``simple'' type-$s$ string operator $B_p^s$, which is also denoted as ${W_s }\left( {\partial p} \right)$. Since the closed string operator $B_p$ does not cross the honeycomb lattice, we can reduce a fattened lattice picture to a linear combination of string-net configurations with only F-moves. As discussed in Sec. \ref{sec.IA}, the F-moves change the fusion sequences of anyons. The algebraic form of F-move is given by a six-index tensor $F$ that explains how the outcomes change under the control of these four anyons. We illustrate this process in Fig. \ref{fig-simple-wilson-seq}\textbf{a}, where the ``controlling'' strings are marked in green and the fusion outcomes are marked in red. 

In Fig. \ref{fig-simple-wilson-seq}\textbf{b} and \textbf{c}, we plot the same process described by $B_p^s$ in the fattened lattice picture and the tailed string-net picture, respectively. 
In Fig. \ref{fig-simple-wilson-seq}\textbf{b}, we show how to change a fattened lattice picture to a superposition of string-net configurations, where the corresponding coefficients and the Dirac kets are omitted for brevity. 
We note that the superposition coefficients in Fig. \ref{fig-simple-wilson-seq}\textbf{b} are equal to that given in Eq. (\ref{eq:Bps-graphic}) despite the different forms. 
We can also recognize this Wilson loop as the process of creating two quasiparticles (tails in Fig. \ref{fig-simple-wilson-seq}\textbf{c} and \textbf{d}) and fusing them later. The blue strings in these two figures represent the ``occupied'' (type-$1$) string \cite{Levin2005String} to show some of the valid string-net configurations in this process for the Fibonacci string-net model.

\begin{figure}[htb]
\includegraphics[width=0.9\linewidth]{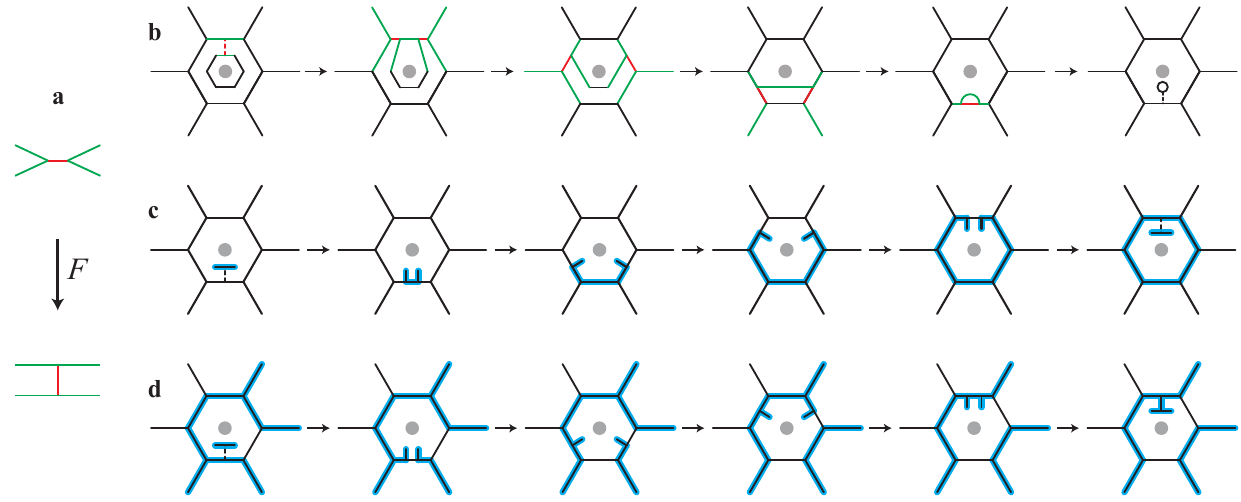}
\caption{\label{fig-simple-wilson-seq} 
Description of the shortest closed simple string operator $W\left( {\partial p} \right)$ in different pictures. \textbf{a}, The F-move on unoriented strings. The green strings are "controlling" string which is unchanged after the F-move. The red strings represent fusion outcomes that change according to the known string types to be fused. \textbf{b}, The transformation from the fattened lattice to the superposition of string-net configurations after implementing the $W\left( {\partial p} \right)$ operator on the string-net wavefunction.
The coefficients describing the linear transformation are omitted. 
\textbf{c} and \textbf{d}, The tailed string-net picture describing the same process as in \textbf{b}. We omit the string connecting the endpoints of the two tails. This picture shows the process of creating two quasiparticles (blue strings) and annihilating them to vacuum more intuitively. 
We use the thick blue lines to represent the type-$1$ strings in the Fibonacci string-net model whose fusion rule is given by Eq. (\ref{eq:supp-fusion-rule-Fibonacci}). 
For clarity, here we present two examples of the valid string-net configurations in \textbf{c} and \textbf{d}, respectively.
The dotted lines represent the strings that are known to be vacuum.}
\end{figure}

\begin{figure}[htb]
\includegraphics[width=0.8\linewidth]{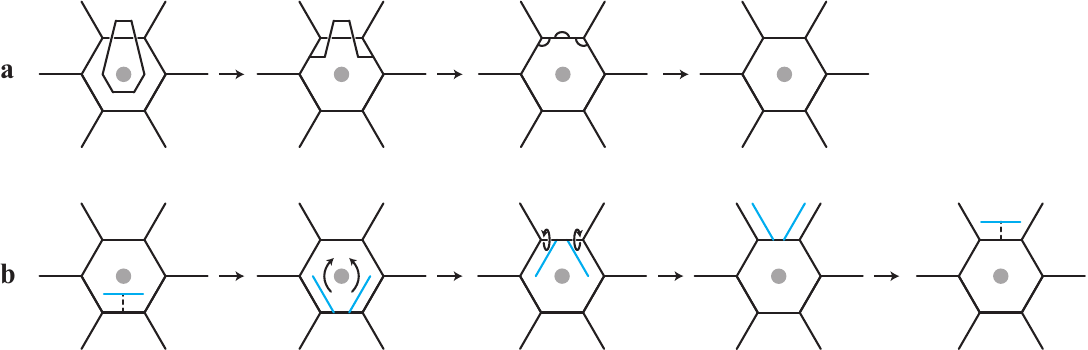}
\caption{\label{fig-simple-wilson-cross} A closed string operator crossing the honeycomb lattice twice. \textbf{a}, Reduction from the fattened lattice picture to a superposition of string-net configurations. \textbf{b}, The tailed string-net picture describing the same process as \textbf{a}. The rotations shown in the third subgraph contain a clockwise R move and an anticlockwise ${{\text R}^{ - 1}}$ move.}
\end{figure}

The reduction of closed string operator crossing different plaquettes in fattened lattice picture to the superposition of string-net configurations requires the relationship characterized by the $\Omega$-matrix, as shown in Eq. (\ref{eq:Omega-relation}), in addition to those given by the F-matrix. We still use the tailed string-net picture which better illustrates this process and is more convenient in the design of quantum circuits. An example of describing the same process is shown in Fig. \ref{fig-simple-wilson-cross}. We note that the $\Omega$-matrix and the R-matrix characterize the same relationship between different configurations since they are both solved from the hexagon equation.

In this work, we use the tailed string-net configuration to better illustrate the position of the electric charge $\tau$ in the absence of the fluxon (for further information about the full dyon spectrum, see Ref. \cite{Hu2018Full}). The electric charge can be defined with an extra $Q_v$ operator on the tail, which is:
\begin{align}\label{eq:Electric-charge}
Q_v
\left| \raisebox{-0.45cm}{\includegraphics[scale=0.35]{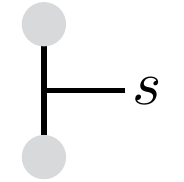}} \right\rangle = \delta_{s,0}
\left| \raisebox{-0.45cm}{\includegraphics[scale=0.35]{eq/electric_charge.pdf}} \right\rangle.
\end{align}
The creation, movement, and fusion of these electric charges have been introduced as the endpoints of the string operators (tails) in the main text. With the F- and R-move, these quasiparticles can be moved to demonstrate their topological spin and mutual statistics. The non-Abelian mutual statistics of Fibonacci anyon provide a promising way to define logical qubits and operations in topological quantum computation, which is demonstrated in our experiment.

\subsection{Closed string operators in the tailed string-net picture}

We have introduced and experimentally demonstrated that we can detect the electric charge with the extra $Q_v$ operator defined in Eq. (\ref{eq:Electric-charge}). We also would like to further introduce an intuitive way to derive the nonlocal closed string operator to detect the quasiparticle excitations, which is known as Ocneanu's tube algebra in TQFT \cite{Schotte202204Quantum}. The definition of these different closed string operators characterizing the same topological charge is important for the error correction scheme in fault-tolerant quantum computing.

\begin{figure}[htb]
\includegraphics[width=1\linewidth]{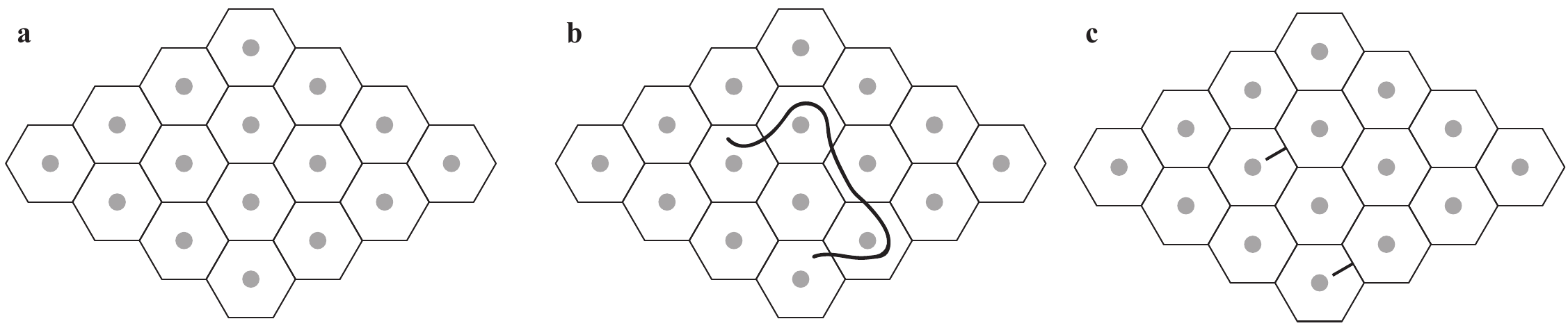}
\caption{\label{fig-tailed-string} The string operator in fattened lattice and the tailed string-net picture. \textbf{a}, The ground state of the original string-net picture. \textbf{b}, An open string operator in the fattened lattice picture. \textbf{c}, The reduced tailed string-net picture corresponding to the string operator of \textbf{b}.}
\end{figure}

In the tailed string-net picture shown in Fig. \ref{fig-tailed-string}\textbf{a}, the action of an open string operator that creates quasiparticle excitations can be directly drawn on it as shown in Fig. \ref{fig-tailed-string}\textbf{b}. This string state can be decomposed to a linear combination of tailed string-net with the geometry shown in Fig. \ref{fig-tailed-string}\textbf{c}. To derive the algebraic form of the fixed-point Hamiltonian, we first consider a closed string operator containing one tail as shown in Fig. \ref{fig-tailed-closed-string}\textbf{a}. We note that this closed string operator is defined on strings far from the tails, whose algebraic form does not change after adding the tails. Since there is no tail in other plaquettes (or all other tails are vacuum strings), this close string operator can be continuously transformed to a smaller one as shown in Fig. \ref{fig-tailed-closed-string}\textbf{b}. According to the hexagon equation of Eq. (\ref{eq:Omega-hexagon}), the closed string operator can be reduced to the form shown in Fig. \ref{fig-tailed-closed-string}\textbf{c}.

\begin{figure}[htb]
\includegraphics[width=1\linewidth]{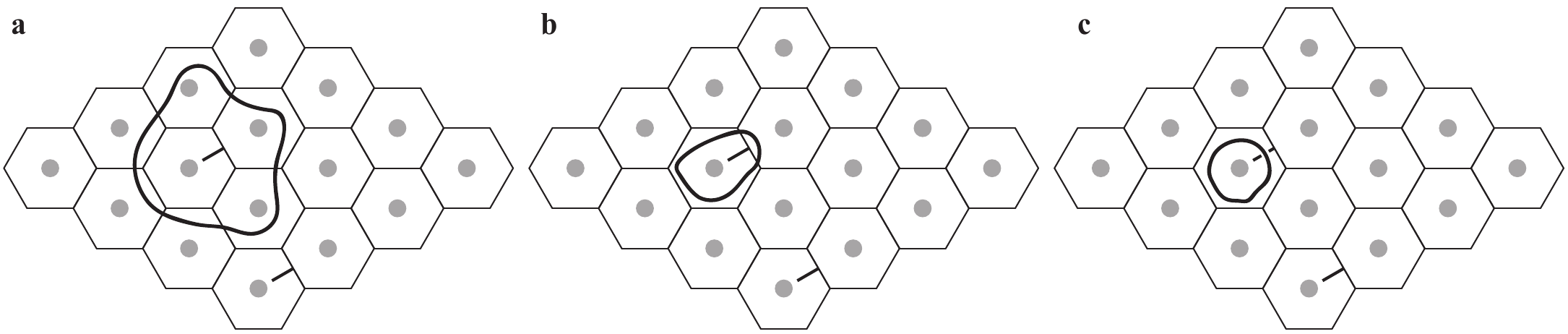}
\caption{\label{fig-tailed-closed-string} Topological equivalent closed string operators. \textbf{a}, A large close string operator containing one tail. Its algebraic form is the same as the one in the original string-net picture. \textbf{b}, A smaller topological equivalent closed string operators. \textbf{c}, The smallest topological equivalent closed string operators.}
\end{figure}

\begin{figure}[htb]
\includegraphics[width=1\linewidth]{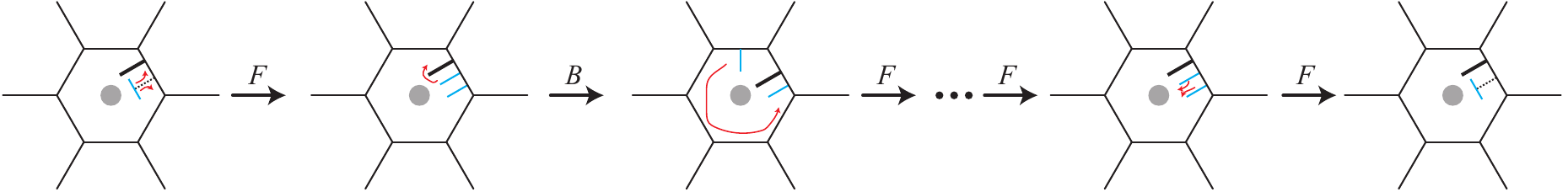}
\caption{\label{fig-tailed-process}A description of simple closed string operator $O_p^x$ in the tailed string-net picture. We can directly derive the algebraic form in Eq. (\ref{eq:Ops-algebra}) according to this figure.}
\end{figure}

To derive the algebraic form of this closed string operator, we rewrite it as a process shown in Fig. \ref{fig-tailed-process}. The process can be described as: (1) create one pair of type-$x$ excitations near the tail; (2) move one of the quasiparticles across the tail with a $B$ operator given in Eq. (\ref{eq:B-tensor}); (3) wind the above quasiparticle around this plaquette; (4) annihilate the two quasiparticles to vacuum. The corresponding algebraic form of this closed string operator $O_p^x$ can be written as:

\begin{align}\label{eq:Ops-algebra}
O_p^x
\stretchleftright[800]{\bigg |}{\adjustbox{valign=m}{\includegraphics[scale=1.3]{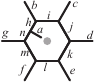}}}{\Bigg\rangle}
= 
\stretchleftright[800]{\bigg |}{\adjustbox{valign=m}{\includegraphics[scale=1.3]{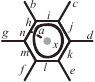}}}{\Bigg\rangle}
= 
\sum\limits_{xopqrstu} {B_{uxo}^{ahn}F_{xop}^{bih}F_{xpq}^{cji}F_{xqr}^{dkj}F_{xrs}^{elk}F_{xst}^{fml}F_{xtu}^{gnm}
\stretchleftright[800]{\bigg |}{\adjustbox{valign=m}{\includegraphics[scale=1.3]{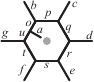}}}{\Bigg\rangle}
}.
\end{align}

This closed string operator can characterize the topological charge of an area. This corresponds to a basic fact that the way to detect a particle is to interact with it. In this double-Fibonacci model, we can detect whether there is a $\tau$ charge by surrounding this area with a $\tau$ anyon.

Similar to the extra $Q_v$ defined in Eq. (\ref{eq:Electric-charge}), we can define the extra $B_p$ operator $O_p$ as:
\begin{align}\label{eq:Ops-algebra-1}
{O_p} = {1 \over D}\sum\limits_s {{d_s}O_p^s}.
\end{align}
It can be easily proved that the $O_p$ is a projector from the fattened lattice picture shown in Fig. \ref{fig-tailed-closed-string}. The ground state is the eigenstate of  $O_p$ with the eigenvalue of $1$, since the ground state does not have tails and the operator $O_p$ is reduced to $B_p$. Meanwhile, the measurement result of $O_p$ on the result state $\left| {{\psi _{\tau \tau }}} \right\rangle $ after implementing a type-$\tau$ string operator is zero. Now we explain this statement with a basic physical intuition: the state with a pair of quasiparticle excitations should be orthogonal to the state with their time reversal counterpart excitations. Considering the ground state on a closed loop with one string, the overlap between states with two kinds of quasiparticle excitations related by the time-reversal symmetry is zero for the self-dual model:
\begin{align}\label{eq:orth-double}
0 = {1 \over {{D^2}}}\sum\limits_{ij} {{d_i}{d_j}\left\langle {\raisebox{-0.3cm}{\includegraphics[scale=1]{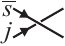}}} \right|\left. {\raisebox{-0.3cm}{\includegraphics[scale=1]{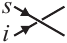}}} \right\rangle } & = {1 \over {{D^2}}}\sum\limits_{ijmn} {{d_i}{d_j}F_{s^*sm}^{ii^*0}F_{\bar s^*\bar sn}^{jj^*0}R_{is}^m\left( {R_{j\bar s}^n} \right)^*\left\langle {\raisebox{-0.3cm}{\includegraphics[scale=1]{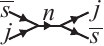}}} \right|\left. {\raisebox{-0.3cm}{\includegraphics[scale=1]{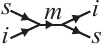}}} \right\rangle } \nonumber \\ &  = \sum\limits_{im} {{{{d_i}{d_m}} \over {{d_s}}}R_{is}^m(R_{i\bar s}^m)^*}.
\end{align}
For the self-dual model with time reversal symmetry, $R_{i\bar s}^m = {\left( {R_{{i^*}s}^m} \right)^{ - 1}} = {\left( {R_{is}^m} \right)^{ - 1}} = (R{_{is}^m)^*}$, and 
\begin{align}\label{eq:iden-reversal}
\left| {\raisebox{-0.3cm}{\includegraphics[scale=1]{eq/phi-tau.pdf}}} \right\rangle  = \sum\limits_m {F_{ssm}^{ii0}} R_{is}^m\left| {\raisebox{-0.3cm}{\includegraphics[scale=1]{eq/phi-tau-m.pdf}}} \right\rangle  = \sum\limits_m {F_{ssm}^{ii0}} {\left( {R_{i\bar s}^m} \right)^{ - 1}}\left| {\raisebox{-0.3cm}{\includegraphics[scale=1]{eq/phi-tau-m.pdf}}} \right\rangle  = \left| {\raisebox{-0.3cm}{\includegraphics[scale=1]{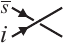}}} \right\rangle .
\end{align}
Thus,
\begin{align}\label{eq:orth-updown}
\left\langle {\raisebox{-0.3cm}{\includegraphics[scale=1]{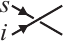}}} \right|\left. {\raisebox{-0.3cm}{\includegraphics[scale=1]{eq/phi-tau.pdf}}} \right\rangle  = 0.
\end{align}
We can directly derive that there is no ``overlap'' between the operator $O_pW_{\tau}$ and ${W_{\tau }}{O_p}$ for the fattened lattice representations of $O_p$ and $W_{\tau}$. Thus, the measurement result is
\begin{align}\label{eq:Op=0}
\left\langle {{\psi _{\tau \tau }}} \right|{O_p}\left| {{\psi _{\tau \tau }}} \right\rangle  = \left\langle \psi  \right|{\left( {{W_\tau }} \right)^\dag }{O_p}{W_\tau }\left| \psi  \right\rangle  = \left\langle \psi  \right|{\left( {{W_\tau }{O_p}} \right)^\dag }{O_p}{W_\tau }\left| \psi  \right\rangle  = 0.
\end{align}
We can conclude that the string operator  $W_{\tau}$ defined in the tailed string-net picture can create the type-$\tau$ excitations where the corresponding $\left\langle O_p \right\rangle  = 0$. The relationship shown in Eq. (\ref{eq:orth-double}) is associated with the braiding nondegeneracy, which is described formally in p67-p87 of Ref. \cite{kitaev2006anyons}.

\subsection{Topological entanglement entropy}

In the string-net picture, strings are connected by the trivalent (and bivalent) points (Fig. \ref{fig-TEE-R}\textbf{a}). To make the boundary more symmetric, we attach two spins (qubits) on each string instead of one, as shown in Fig. \ref{fig-TEE-R}\textbf{b}. These two qubits represent the same type of directly connecting strings. The entanglement entropy in the region $R$ of the string net model can be calculated as \cite{Levin2006Detecting}
\begin{align}\label{eq:TEE-R}
{S_R} =  - j\log \left( D \right) - n\sum\limits_i^N {{{d_k^2} \over D}\log \left( {{{{d_k}} \over D}} \right)},
\end{align}
where $n$ is the number of strings along $\partial R$, $j$ is the number of disconnected boundary curves in $\partial R$ \cite{Levin2006Detecting}, and $S_R$ is the von Neumann entropy of the reduced density matrix ${\rho _R}$. For example, region A shown in Fig. \ref{fig-TEE-R}\textbf{b} has $n=3$ strings along $\partial A$ and is a simply connected region with $j=1$. In contrast, region D shown in Fig. \ref{fig-TEE-R}\textbf{c} has two disconnected boundary curves and twelve strings along the boundary $\partial D$. According to Eq. (\ref{eq:TEE-R}), one way to calculate the topological entanglement entropy is \cite{Kitaev2006Topological}
\begin{align}\label{eq:TEE-ABC}
{S_{topo}} \equiv {S_A} + {S_B} + {S_C} - {S_{AB}} - {S_{BC}} - {S_{AC}} + {S_{ABC}}.
\end{align}

\begin{figure}[htb]
\includegraphics[width=0.7\linewidth]{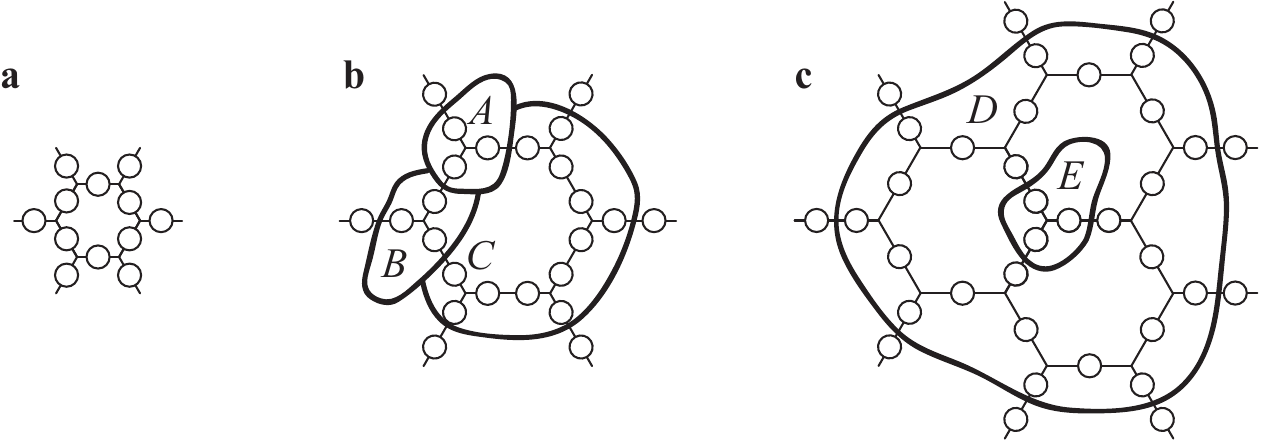}
\caption{\label{fig-TEE-R} Different regions on the lattice. \textbf{a}, One plaquette with twelve strings (qubits). \textbf{b}, The same lattice as \textbf{a} where each string is attached with two qubits. \textbf{c}, Three plaquettes where we define a region D with two disconnected boundary curves.}
\end{figure}

In the string-net model, the operation of copying strings on the boundary is to diagonalize the reduced density matrix and extract the information of the quantum dimensions. In fact, the derivation of Eq. (\ref{eq:TEE-R}) is a diagonalization process of the reduced density matrix. We can deform the geometry of the string-net configurations in the region $R$ with F-moves as shown in Fig. \ref{fig-TEE-F}. This process corresponds to a unitary operation that can be explicitly given, though complicated. Luckily, we do not need to perform such a complicated quantum circuit if we copy the strings on the boundary. This copy operation localizes the unitary inside the region $R$, which does not change the entanglement spectrum of the reduced density matrix before and after the deformation. 

\begin{figure}[htb]
\includegraphics[width=0.9\linewidth]{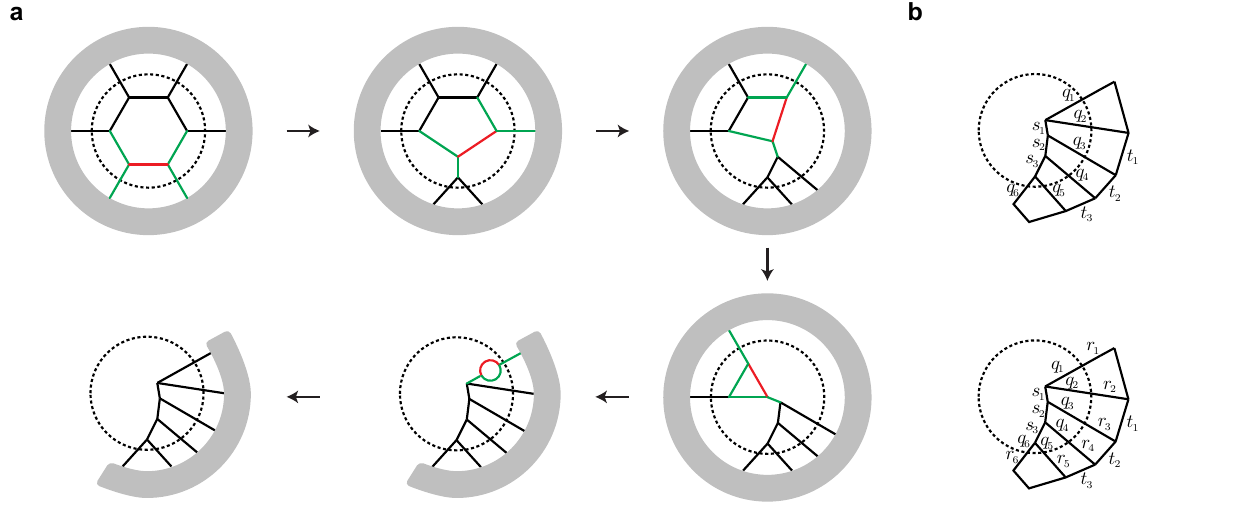}
\caption{\label{fig-TEE-F}Diagonalization of the reduced density matrix. After copying the strings on the boundary, the process of deforming the string-net configurations in the region $R$ to the final geometry is a unitary process to diagonalize the reduced density matrix ${\rho _R}$. After the deformation, the reduced density matrix is a diagonal matrix whose elements are determined by the types of strings on the boundary. The measurement of $S_\text{topo}$ does not actually perform the unitary consisting of the F-moves, since the entanglement spectrum does not change under a unitary operation.}
\end{figure}

In the penultimate step, we notice the linear relationship:
\begin{align}\label{eq:TEE-amplitude}
\left| {\raisebox{-0.45cm}{\includegraphics[scale=1.3]{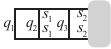}}} \right\rangle  = F_{{q_1}{s_1}0}^{{s_1}{q_1}{q_2}}\left| {\raisebox{-0.45cm}{\includegraphics[scale=1.3]{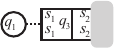}}} \right\rangle  = {{{v_{{q_2}}}} \over {{v_{{q_1}}}{v_{{s_1}}}}}\left| {\raisebox{-0.45cm}{\includegraphics[scale=1.3]{eq/tapole-tee.pdf}}} \right\rangle  = {{{v_{{q_1}}}{v_{{q_2}}}} \over {{v_{{s_1}}}}}\left| {\raisebox{-0.45cm}{\includegraphics[scale=1.3]{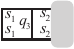}}} \right\rangle .
\end{align}
Thus we can explicitly give the wavefunction as (without normalization),
\begin{align}\label{eq:phi-boundary}
\sum\limits_{\left\{ {q,s,t} \right\}} {{a_{\left\{ q \right\}}}{\delta _{\left\{ s \right\},\left\{ t \right\}}}\left| {{q_1},{q_2}, \cdots } \right\rangle \left| {{s_1},{s_2}, \cdots } \right\rangle \left| {{t_1},{t_2}, \cdots } \right\rangle },
\end{align}
where the coefficients can be strictly calculated as ${a_{\left\{ q \right\}}} = \prod\limits_{\left\{ q \right\}} {{v_q}} $. In the simply connected case with $j=1$, one can verify $ - \sum\limits_{\left\{ {q,s,t} \right\}} {{{a_{\left\{ q \right\}}^2} \over Z}\log \left( {{{a_{\left\{ q \right\}}^2} \over Z}} \right)}  =  - \log \left( D \right) - n\sum\limits_i^N {{{d_k^2} \over D}\log \left( {{{{d_k}} \over D}} \right)} $, where $Z = \sum\limits_{\left\{ {q,s} \right\}} {a_{\left\{ q \right\}}^2} $ is the normalization factor. This result indicates that we can implement the quantum circuits to deform the geometry of the string-net configurations, and measure the probabilities of ${\left| {q_{_1}},{q_2}, \cdots \right\rangle }$ to obtain the information of quantum dimensions. However, the implementation of the transforming quantum circuits is complicated. An alternative way is to copy the strings on the boundaries of the region $R$, the state after the deformation becomes,
\begin{align}\label{eq:phi-boundary-copied}
\sum\limits_{\left\{ {q,r,s,t} \right\}} {{a_{\left\{ q \right\}}}{\delta _{\left\{ q \right\},\left\{ r \right\}}}{\delta _{\left\{ s \right\},\left\{ t \right\}}}\left| {{q_1},{q_2}, \cdots } \right\rangle \left| {{r_1},{r_2}, \cdots } \right\rangle \left| {{s_1},{s_2}, \cdots } \right\rangle \left| {{t_1},{t_2}, \cdots } \right\rangle },
\end{align}
where the copy operation introduces a Delta function ${{\delta _{\left\{ q \right\},\left\{ r \right\}}}}$ which is crucial in the derivation of Eq. (\ref{eq:TEE-R}).
The reduced density matrix of the region $R$ can be calculated as (without normalization),
\begin{align}\label{eq:density-R-copied}
& \mathop {Tr}\limits_{\left\{ {r,t,r',t'} \right\}} \left( {\sum\limits_{\left\{ {q,r,s,t} \right\}} {{a_{\left\{ q \right\}}}{\delta _{\left\{ q \right\},\left\{ r \right\}}}{\delta _{\left\{ s \right\},\left\{ t \right\}}}\left| q \right\rangle \left| r \right\rangle \left| s \right\rangle \left| t \right\rangle } \sum\limits_{\{ q',r',s',t'\} } {{a_{\left\{ {q'} \right\}}}{\delta _{\left\{ {q'} \right\},\left\{ {r'} \right\}}}{\delta _{\left\{ {s'} \right\},\left\{ {t'} \right\}}}\left\langle {q'} \right|\left\langle {r'} \right|\left\langle {s'} \right|\left\langle {t'} \right|} } \right) \nonumber \\ &
= \sum\limits_{\left\{ {q,r,s,t} \right\}} {\sum\limits_{\{ q',r',s',t'\} } {\sum\limits_{\left\{ {r'',t''} \right\}} {{a_{\left\{ {q'} \right\}}}{\delta _{\left\{ {q'} \right\},\left\{ {r'} \right\}}}{\delta _{\left\{ {s'} \right\},\left\{ {t'} \right\}}}\left| q \right\rangle \left| r \right\rangle \left| s \right\rangle \left| t \right\rangle \left\langle {q'} \right|\left\langle {r'} \right|\left\langle {s'} \right|\left\langle {t'} \right|\left( {\left| {r''} \right\rangle \left\langle {r''} \right| \otimes \left| {t''} \right\rangle \left\langle {t''} \right|} \right)} } } \nonumber \\ &
= \sum\limits_{\left\{ {q,r,s,t} \right\}} {\sum\limits_{\{ q',r',s',t'\} } {\sum\limits_{\left\{ {r'',t''} \right\}} {{a_{\left\{ q \right\}}}{\delta _{\left\{ q \right\},\left\{ r \right\}}}{\delta _{\left\{ s \right\},\left\{ t \right\}}}{a_{\left\{ {q'} \right\}}}{\delta _{\left\{ {q'} \right\},\left\{ {r'} \right\}}}{\delta _{\left\{ {s'} \right\},\left\{ {t'} \right\}}}\left| q \right\rangle \left\langle {r''} \right.\left| r \right\rangle \left| s \right\rangle \left\langle {t''} \right.\left| t \right\rangle \left\langle {q'} \right|\left\langle {r'} \right|\left. {r''} \right\rangle \left\langle {s'} \right|\left\langle {t'} \right|\left. {t''} \right\rangle } } }  \nonumber \\ &
= \sum\limits_{\left\{ {q,r,s,t} \right\}} {\sum\limits_{\{ q',r',s',t'\} } {\sum\limits_{\left\{ {r'',t''} \right\}} {{a_{\left\{ q \right\}}}{a_{\left\{ {q'} \right\}}}\left( {{\delta _{\left\{ q \right\},\left\{ r \right\}}}{\delta _{\left\{ r \right\},\left\{ {r''} \right\}}}{\delta _{\left\{ {r'} \right\},\left\{ {r''} \right\}}}{\delta _{\left\{ {q'} \right\},\left\{ {r'} \right\}}}} \right)\left( {{\delta _{\left\{ s \right\},\left\{ t \right\}}}{\delta _{\left\{ t \right\},\left\{ {t''} \right\}}}{\delta _{\left\{ {t'} \right\},\left\{ {t''} \right\}}}{\delta _{\left\{ {s'} \right\},\left\{ {t'} \right\}}}} \right)\left| q \right\rangle \left| s \right\rangle \left\langle {q'} \right|\left\langle {s'} \right|} } }  \nonumber \\ &
= \sum\limits_{\left\{ {q,s} \right\}} {{a_{\left\{ q \right\}}}{a_{\left\{ q \right\}}}\left| q \right\rangle \left| s \right\rangle \left\langle q \right|\left\langle s \right|}.
\end{align}
This is a diagonal matrix in the bases of $\left| q \right\rangle \left| r \right\rangle$, where, for example, $\left| q \right\rangle $ is the abbreviation of $\left| {{q_1},{q_2}, \cdots } \right\rangle$. The transformation of string-net configurations can be individually achieved inside and outside the region $R$, which also does not change the entanglement spectrum. The entanglement entropy of original reduced density matrix where the strings on the boundaries are copied is also ${S_R} =  - j\log \left( D \right) - n\sum\limits_i^N {{{d_k^2} \over D}\log \left( {{{{d_k}} \over D}} \right)}$. In short, the duplication of strings on the boundary is based on that the reduced density matrix of state $\sum\limits_i {{c_i}\left| {{\phi _i}} \right\rangle \left| {{\phi _i}} \right\rangle \left| {{\psi _i}} \right\rangle } $ is diagonal under the bases of ${\left| {{\phi _i}} \right\rangle }$, which is the case of Eq. (\ref{eq:density-R-copied}).

We would like to intuitively describe the difference between the string-net picture and the toric code \cite{Satzinger2021Realizing} when measuring the $S_\text{topo}$. In the toric code, two adjacent qubits are connected by a string, which can be taken as a bivalent point. Thus, two regions can be defined as each containing one of the two qubits. However, the strings (qubits) in the string-net are connected by the trivalent vertices. If we do not copy the qubits on the boundaries, the qubit representing the string crossing the boundary can only be assigned to one region, which makes Eq.(\ref{eq:density-R-copied}) invalid.

\section{Circuit Design}
\subsection{Formalism of the Fibonacci string-net model}
In this work, we created the ground state of the Fibonacci string-net model and simulated the braiding of associated quasiparticle excitations, whose nontrivial fusion rule is given in Eq. (\ref{eq:supp-fusion-rule-Fibonacci}). Referring to Eq. (\ref{eq:Qv-graphic}), the $Q_v$ operator of the Fibonacci string-net model can be explicitly given as,
\begin{align}\label{eq:Qv-Fibonacci}
{Q_v}\left| {\raisebox{-0.3cm}{\includegraphics[scale=1.0]{eq/vertex-Qv.pdf}}} \right\rangle  = {\delta _{ijk}}\left| {\raisebox{-0.3cm}{\includegraphics[scale=1.0]{eq/vertex-Qv.pdf}}} \right\rangle \text{, where } {\delta _{ijk}} = 
\begin{cases}
1, & \text{if } ijk = 000,011,101,110,111,\\
0,    & \text{otherwise}.
\end{cases}
\end{align}
We note that ${\delta _{000}} = 1$ is a trivial valid fusion rule. Another trivial fusion rule is that the null string can be arbitrarily added and removed. Thus, the bivalent vertex of an anyon and its dual are always valid. This corresponds to that ${\delta _{011}}$, ${\delta _{101}}$, and ${\delta _{110}}$ are valid for Fibonacci anyons, which is self-dual ${1^*} = 1$. Two Fibonacci anyons may also fuse to another Fibonacci anyon, corresponding to the valid ${\delta _{111}} = 1$.

The $B_p$ operator is given by the 6-index $F$ tensor, which can be solved from Eq. (\ref{eq:F-constraints}). Most elements can be easily calculated with the constraints of physicality and normalization. For example, $F_{\tau \tau \tau}^{{\bf{111}} } = 1$ and $F_{\tau \tau {\bf{1}}}^{{\bf{111}} } = 0$ because of the normalization of Eq. (\ref{eq:strnet-normalization}). There are only four elements of the F-matrix that do not equal 0 or 1, which are $F_{\tau \tau {\bf{1}}}^{\tau \tau {\bf{1}}} = {1 \over \phi }$, $F_{\tau \tau \tau }^{\tau \tau {\bf{1}}} = {1 \over {\sqrt \phi  }}$, $F_{\tau \tau {\bf{1}}}^{\tau \tau \tau } = {1 \over {\sqrt \phi  }}$, and $F_{\tau \tau \tau }^{\tau \tau \tau } =  - {1 \over \phi }$. We denote these nontrivial elements as a unitary matrix ${U_F} = \left({\begin{matrix} {1/\phi} & {1/\sqrt \phi } \\ {1/\sqrt \phi } &{-1/\phi} \end{matrix}}\right)$. Here $\phi$ is the quantum dimension of Fibonacci anyons, which is ${d_\tau } = {{1 + \sqrt 5 } \over 2}$ solved from Eq. (\ref{eq:D-quantum-rules}) and Eq. (\ref{eq:supp-fusion-rule-Fibonacci}). 
Similarly, $R_{\tau \tau }^{\bf{1}} = {e^{ - 4\pi i/5}}$, and $R_{\tau \tau }^\tau  = {e^{3\pi i/5}}$, which are solved from the hexagon equation shown in Eq. (\ref{eq:Omega-hexagon}) and Fig. \ref{fig-hexagon-equation} and can be denoted collectively as ${U_R} = \left({\begin{matrix} {e^{ - 4\pi i/5}} & {0} \\ {0} &{e^{3\pi i/5}} \end{matrix}}\right)$ . The conjugate of the R-matrix is also a valid solution, which describes the time reversal counterpart ${\bar \tau }$ of the Fibonacci anyon $\tau$.

\subsection{Circuits for the F- and R-moves of Fibonacci anyon}

\begin{figure}[htb]
\includegraphics[width=1\linewidth]{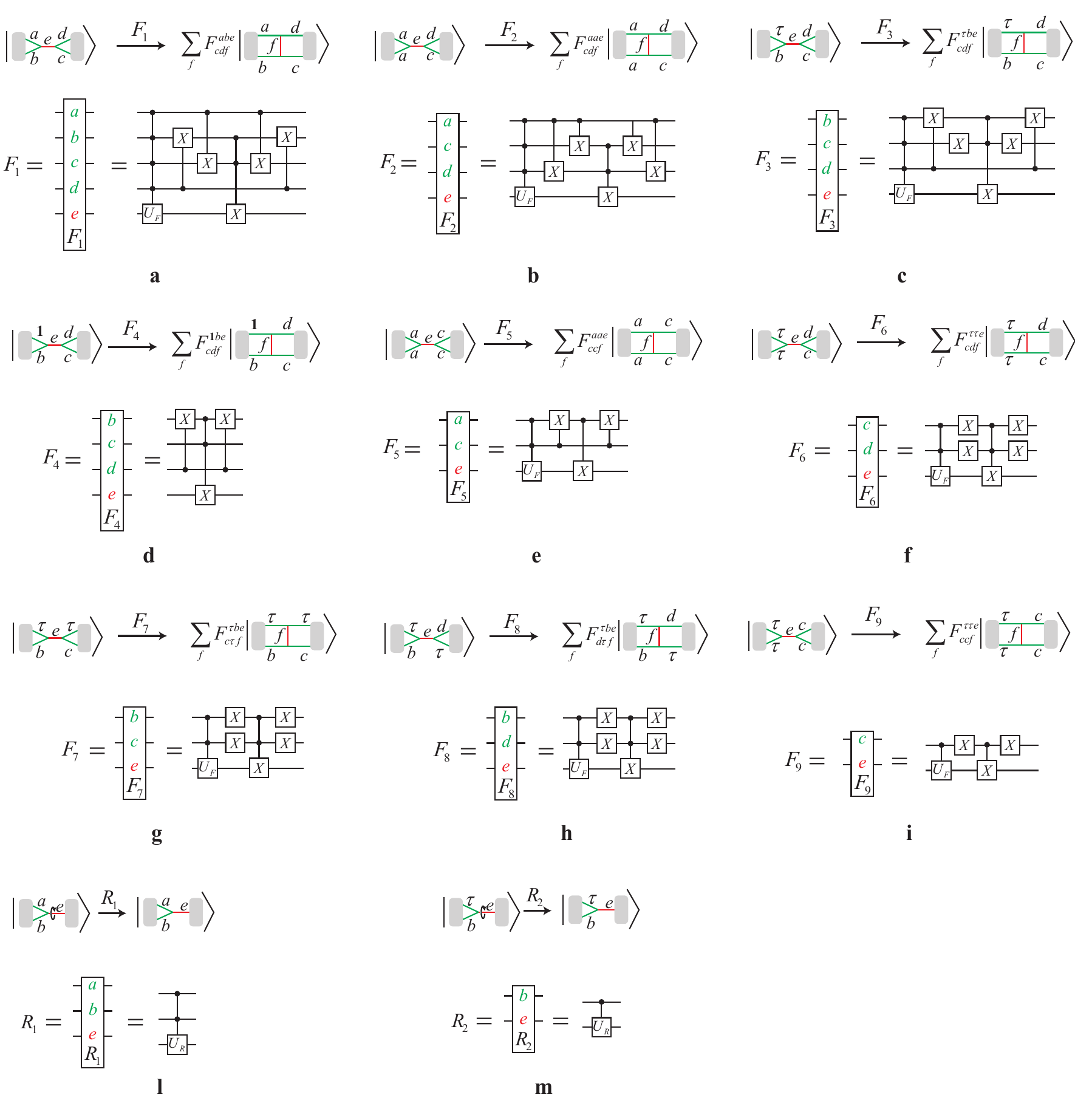}
\caption{\label{fig-f-circuits} The quantum circuits corresponding to the F- and R-moves under different conditions. The green strings (qubits) are unchanged after the different types of F-moves, and the red strings (qubits) corresponding to $e$ are changed to a new string type $f$. The R-moves are realized by implementing a unitary gate on the red qubits controlled by the green qubits. \textbf{a}, The circuit to implement the full F-move. \textbf{b}-\textbf{i}, The circuit to implement the full F-move when $a=b$ (\textbf{b}), $a=\tau$ (\textbf{c}), $a={\bf 1}$ (\textbf{d}), $a=b$ and $c=d$ (\textbf{e}), $a=b=\tau$ (\textbf{f}), $a=d=\tau$ (\textbf{g}), $a=c=\tau$ (\textbf{h}), and $a=b=\tau$ and $c=d$ (i).
\textbf{l} and \textbf{m}, The circuit to implement the R-move when $a=d=\tau$ (\textbf{l}) and $a=\tau$ (\textbf{m}).}
\end{figure}

As shown in Eq. (\ref{eq:supp-local-constraints-f}) and Fig. \ref{fig-simple-wilson-seq}\textbf{a}, the F-move acts on five strings and only changes the type of one string. In the quantum circuit scheme, it is a unitary gate that acts on one qubit controlled by four qubits. The corresponding quantum circuit is shown in Fig. \ref{fig-f-circuits}\textbf{a}. We do not need to perform this full circuit when we have some prior information about the string types for the F-move. For example, if the F-move is described by $F_{cdf}^{aae}$, the first two strings can be represented by one qubit and the corresponding circuit is implemented on four qubits as shown in Fig. \ref{fig-f-circuits}\textbf{b}. The quantum circuits for other common situations are denoted as ${F_{3 - 9}}$ and shown in Fig. \ref{fig-f-circuits}\textbf{c}-\textbf{i}. The quantum circuit corresponding to the R move is much simpler as shown in Fig. \ref{fig-f-circuits}\textbf{l}. When one string is known to be $\tau$, the corresponding quantum circuit is reduced to a two-qubit controlled-$U_R$ gate as shown in Fig. \ref{fig-f-circuits}\textbf{m}.
In our work, the physical qubits are located on the vertices of a square lattice, where two-qubit gates can be performed on two nearest neighboring qubits. 
We tilt the honeycomb string-net lattice 30 degrees to fit the geometry of the physical qubits, as shown in Fig. 1\textbf{b} of the main text. 
In Fig. 2\textbf{b} of the main text, we show the quantum circuits of the two F-moves which are used in preparing the ground state.
The circuits are further compiled with a variational method provided by CPFlow \cite{Nemkov2023efficient} to fit the layout topology of the device.
Similarly, the quantum circuits of other F-moves shown in Fig. \ref{fig-f-circuits} can also be decomposed into the form suitable for our chips, which enables us to experimentally measure the quantum dimension of the Fibonacci anyon from its mutual statistics.

\subsection{"Copy" and "discard" qubits}

In the digital quantum simulation of the string-net model, we will frequently copy and delete strings. Taking the creation process shown in Extended Data Fig 3\textbf{a} in the main text as an example, we need one qubit to "copy" the type of the bottom string at first. We also need to implement the copy operation as the first step to measure the topological entanglement entropy. 
To simulate the second local rule in Eq. 5(b) in the main text which deletes closed string loops, the discard operation is required.
These operations can be regarded as entangling and disentangling ancillary qubits. For example, copying one string to another is performed by adding an ancillary qubit of state $\left| 0 \right\rangle $, implementing an X gate on this ancillary qubit controlled by the qubit that carries the information of the string to be copied. Similarly, we can discard the qubits that carry redundant information. Taking the state ${{a\left| {00} \right\rangle  + b\left| {11} \right\rangle } \over {\sqrt {a{a^*} + b{b^*}} }}$ as an example, where $a$ and $b$ are arbitrary complex numbers (not zero at the same time), we can disentangle the second qubit by implementing an X gate on it controlled by the first qubit. The processes of copying and discarding qubits are shown in Fig. \ref{fig-copy-discard}. 

\begin{figure}[htb]
\includegraphics[width=0.7\linewidth]{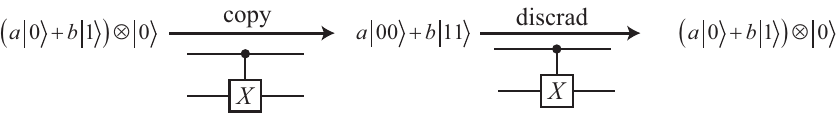}
\caption{\label{fig-copy-discard} The process of copying and discarding qubits. The normalization factor $1\over {\sqrt {a{a^*} + b{b^*}} }$ is omitted for brevity. Initially, the information of string types is stored in the first qubit (the top line). Then a controlled-X (CX) gate copies this information to the second qubit (the bottom line). On the other hand, a CX gate implemented on two qubits which encode the same information, that is, they are on the state ${{a\left| {00} \right\rangle  + b\left| {11} \right\rangle }}$, can reset the second qubit to the state ${\left| 0 \right\rangle }$ which is disentangled from the system.}
\end{figure}

\subsection{Measuring the ${Q_v}$ and $B_p$ operators}

In the main text, we have shown that we can first prepare the ground state of isolate loops and then connect them to the desired geometry with F-moves. After preparing the ground state $\Phi$, we need to verify it by measuring all the terms in the Hamiltonian. Ref. \cite{Bonesteel2012Quantum} proposes detailed circuits to projectively measure the $Q_v$ and the deformed $B_p$ operators with ancillary qubits. Executing these circuits will introduce extra errors in the measurement stage due to limited gate fidelity and coherence time. We use an alternative method of decomposing the $Q_v$ and $B_p$ operators in the Pauli bases, which avoids the error of two-qubit gates in the measurement stage.
The $Q_v$ operators correspond to the fusion rule of Fibonacci anyons, where the phase of the quantum state does not contribute. Thus, we only need to measure three (or two for bivalent point) qubits in the Pauli-Z basis for each $Q_v$. The $B_p$ operator is a projector on twelve qubits, and is decomposed to a sum of {99328} Pauli strings of length 12, which can be measured under {290} Pauli bases.

\subsection{Circuits to detect the braiding statistics}

\begin{figure}[htb]
\includegraphics[width=1\linewidth]{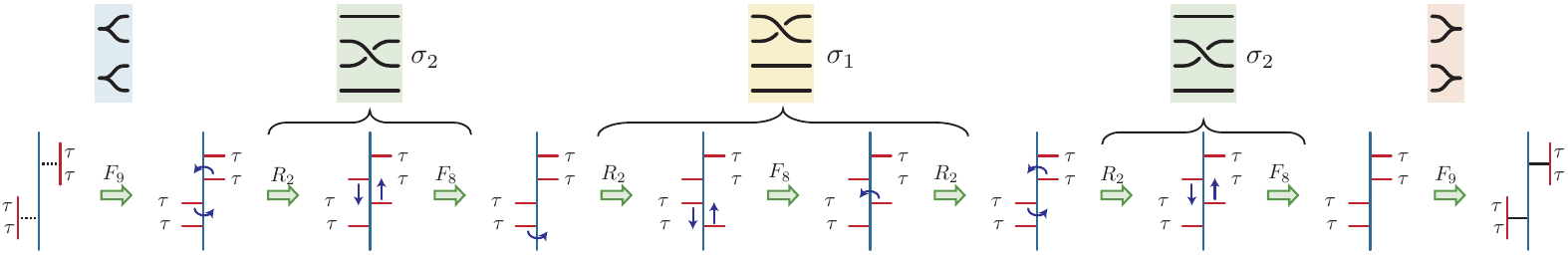}
\caption{\label{fig-detail-all-braid-seq} Detailed process of the longest braiding sequence ${\sigma _2}{\sigma _1}{\sigma _2}$ shown in the main text in the string-net picture. We create two pairs of Fibonacci anyons from vacuum (the dotted lines) with the $F_9$ circuit. The following $\sigma_2$ is performed by implementing two $R_2$ circuits and one $F_8$ circuit, and the $\sigma_1$ is performed by implementing one $R_2$ circuit, one $F_8$ circuit, and another $R_2$ circuit. The circuits corresponding to specific F- and R-moves are shown in Fig. \ref{fig-f-circuits}.}
\end{figure}

As an illustrating example, here we give a detailed description of the longest braiding sequence ${\sigma _2}{\sigma _1}{\sigma _2}$ shown in the main text. As shown in Fig. \ref{fig-detail-all-braid-seq}, we first create two pairs of Fibonacci anyons by adding two open type-$\tau$ strings to the string-net picture and directly connect them to the middle blue string with the $F_9$ circuit. Then we perform the desired braiding by moving the position of Fibonacci anyons (red strings) with F- and R-moves. In the last step, two pairs of Fibonacci anyons are detached from the middle string with another $F_9$ circuit, where we measure the black strings that connect them to the middle string to obtain the fusion results. After the braiding sequence ${\sigma _1}{\sigma _2}$, the corresponding logical state is an eigenstate of the braiding operator $\sigma_2$, which is explained by the Yang-Baxter equation as shown in Eq. (\ref{eq:sigma2-eigen}). This property is verified by our experimental results shown in Fig. 4 of the main text where implementing one more $\sigma_2$ does not change the fusion results.

\section{Experiment}
\label{app:experiment1}

\subsection{Device information}
Our experiments are performed on a quantum processor with frequency-tunable transmon qubits encapsulated in a $11\times 11$ square lattice and tunable couplers between adjacent qubits, with the overall design similar to that in Ref. ~\cite{Xu2023_digital}. The maximum resonance frequencies of the qubit and coupler are around 4.5
GHz and 9.0 GHz, respectively. The effective coupling strength between two neighboring qubits can be dynamically tuned up to -25 MHz. As shown in the Extended Data Fig.~1 of the main text, we use 27 qubits to construct the honeycomb lattice used in this experiment, with their properties shown in Fig.~\ref{fig-property_sq}. The idle frequencies of the 27 qubits where single-qubit gates are applied are shown in Fig.~\ref{fig-property_sq}a.
The relaxation times and the Hahn echo dephasing times measured at the idle frequency of each qubit are also shown in Fig.~\ref{fig-property_sq}b,c, with median values of $T_1 = 117 \mu s$ and $T_2 = 20 \mu s$, respectively. Each qubit can be measured in the computational basis using the dispersive measurement scheme, and the readout fidelities of all the qubits are summarized in Fig.~\ref{fig-property_sq}e,f, which are used to mitigate readout errors (see Sec. \ref{sec-readout_error_mitigation}). 

\subsection{Experimental circuit}
The experimental circuit for preparing the ground state is shown in Fig.~\ref{circuit_ground_state} and Fig.~\ref{circuit_ground_state_2} (see Data availability for the explicit circuits of other experiments in this work), which contains layers of single- and two-qubit gates with different patterns (Fig.~\ref{fig-property_tq}b). We realize arbitrary single-qubit gates by combining XY rotations and Z rotations. Single-qubit XY rotations are implemented by applying 20-ns long microwave pulses with DRAG pulse shaping~\cite{PhysRevLett.119.180511} at qubits' respective idle frequencies, and single-qubit Z rotations are realized by the virtual-Z gate scheme~\cite{PhysRevA.96.022330}. We realize two-qubit CZ gates by bringing $|11\rangle$ and $|02\rangle$ (or $|20\rangle$) of the qubit pairs in near resonance at the respective interaction frequencies and activating the coupling for a specific time of 30 ns, with the calibration procedure detailed in Ref.~\cite{quantum_adversarial_learning}. Below we list two key steps for compiling and executing the experimental circuit.

\subsubsection{Circuit transpilation}
For a given target circuit as described in the main text, we use CPFlow \cite{Nemkov2023efficient} and Qiskit \cite{Qiskit} to reduce its depth and recompile it with single- and two-qubit gates selected from the gate set \{$U(\theta, \varphi, \lambda)$, CZ\}, where $U(\theta, \varphi, \lambda)$ denotes a parameterized single-qubit gate with the following matrix form
\begin{align}\label{eq:u-gate}
U(\theta, \varphi, \lambda)=\left( {\begin{matrix} \cos{\frac{\theta}{2}}&-e^{i\lambda}\sin{\frac{\theta}{2}} \\ e^{i\varphi}\sin{\frac{\theta}{2}} & e^{i(\varphi+\lambda)}\cos{\frac{\theta}{2}} \end{matrix}} \right),
\end{align}
and CZ denotes the two-qubit CZ gate that fits the connectivity topology of our device. In practice, each single-qubit gate is further decomposed into a virtual Z rotation followed by an XY rotation. 
{For each compiled experimental circuit, we further compact it before execution to reduce the impact of dephasing noise.}
We also avoid performing single- and two-qubit gates simultaneously {in the same layer}. As a result, the optimized circuit consists of a staggered arrangement of single- and two-qubit gate layers. During the execution of the circuit, we apply CPMG sequences to further minimize the impact of dephasing, which is realized by applying two Y gates (rotation around the y-axis by $\pi$ angle) on qubits whose successive idle layers span more than six single-qubit gate layers.

\subsubsection{Gate optimization and benchmarking}
{The experimental circuit requires multiple gates in the same layer to be implemented in parallel.} 
In practice, we find that the performances of the parallel single-qubit idle/XY gates and two-qubit CZ gates depend strongly on the idle frequencies and interaction frequencies, which we refer to collectively as gate parameters.
Learning the optimal gate parameters is essential for the successful execution of the experimental circuit, which is also challenging due to the existence of various experimental constraints, system drifts, and the exponentially large search space.
In this work, we follow the idea of the Snake optimizer~\cite{klimov2020snake}, i.e., representing the optimization objects by graphs and reducing the complex high-dimensional optimization problems into multiple simpler lower-dimensional problems, to optimize the gate parameters.

\begin{figure}[htb]
\includegraphics[width=1\linewidth]{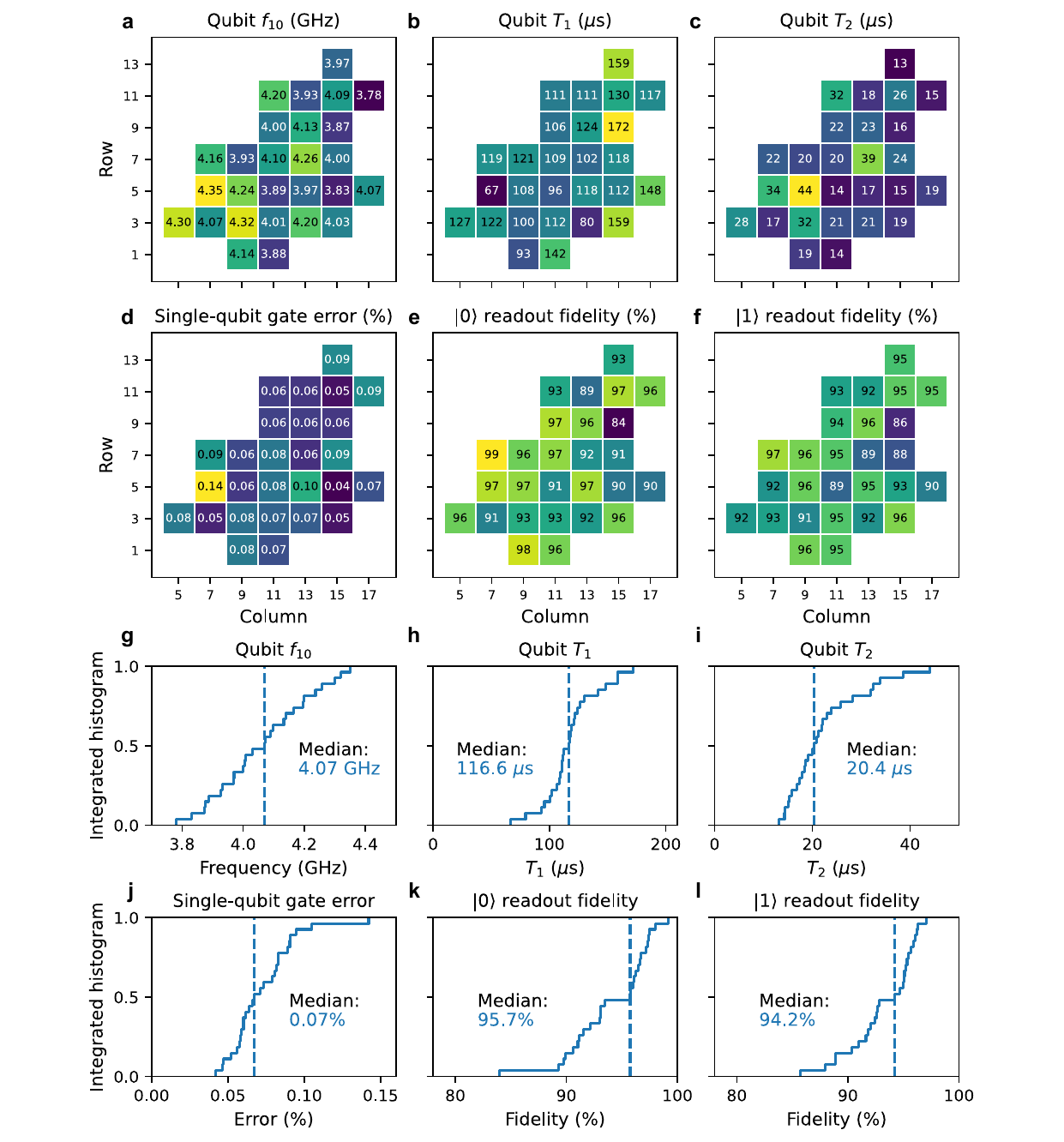}
\caption{\label{fig-property_sq}\textbf{Heatmaps of various single-qubit parameters and corresponding integrated histograms.} \textbf{a}, Qubit idle frequency. \textbf{b}, Qubit relaxation time measured at the idle frequency. \textbf{c}, Qubit dephasing time measured using Hahn echo sequence. \textbf{d}, Pauli error of simultaneous single-qubit gate. \textbf{e}, Readout fidelity of the qubit $|0\rangle$ state. \textbf{f}, Readout fidelity of the qubit $|1\rangle$ state. \textbf{g-l}, Corresponding integrated histograms. Dashed lines indicate the median values.}
\end{figure} 

\begin{figure}[htb]
\includegraphics[width=0.9\linewidth]{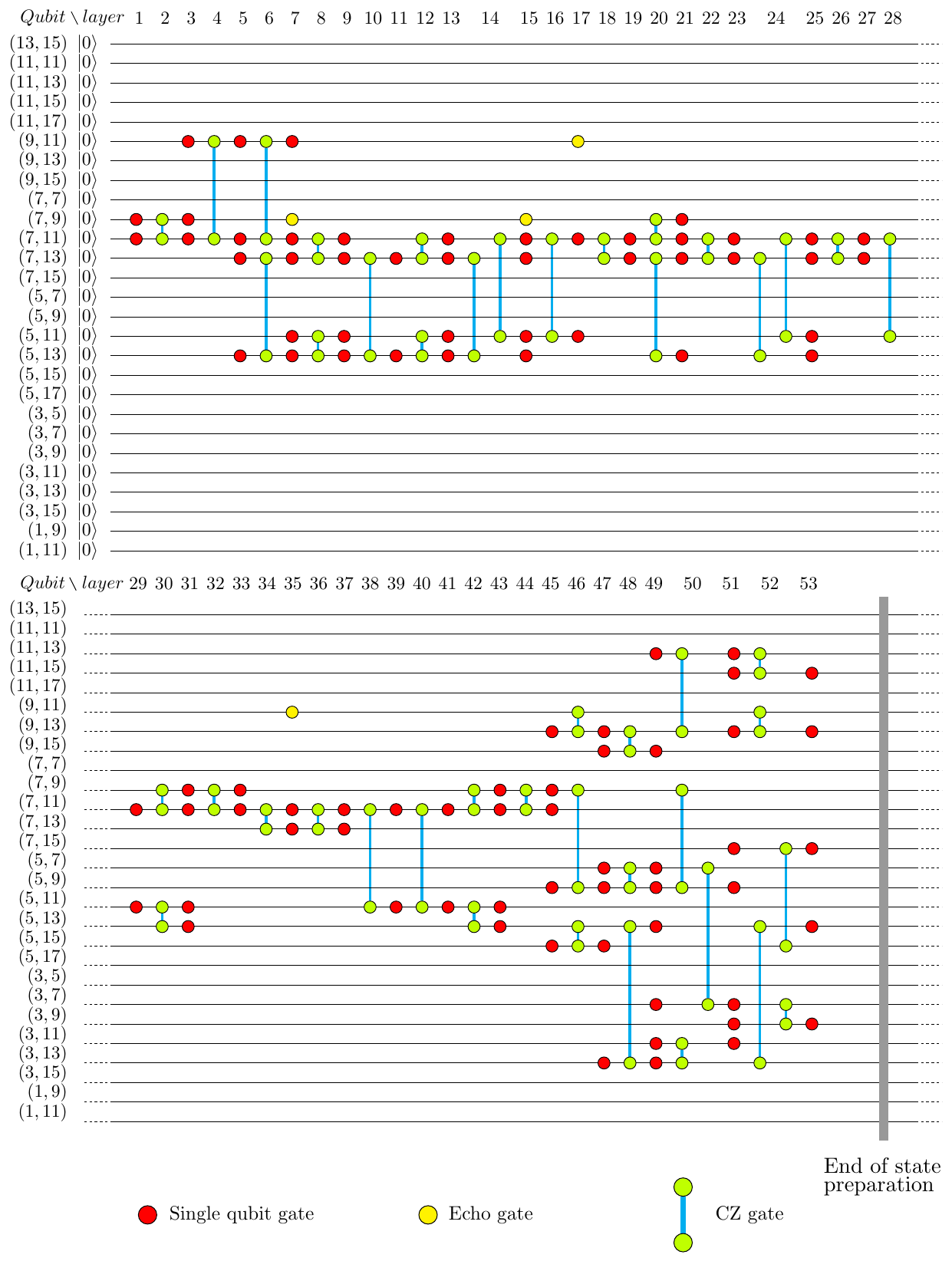}
\caption{\label{circuit_ground_state}\textbf{ Experimental circuit for preparing the ground state of the Fibonacci string-net model}, which is obtained by variationally optimizing the original circuit shown in the main text Fig.~2\textbf{a} targeting the ideal state. The whole circuit contains 53 layers, with each layer containing only single- or two-qubit gates that are applied simultaneously. Each qubit is labeled by (row index, column index) as shown in the Extended Data Fig.~1 of the main text.
}
\end{figure}

\begin{sidewaysfigure}[htb]
\centering
\includegraphics[width=1\linewidth]{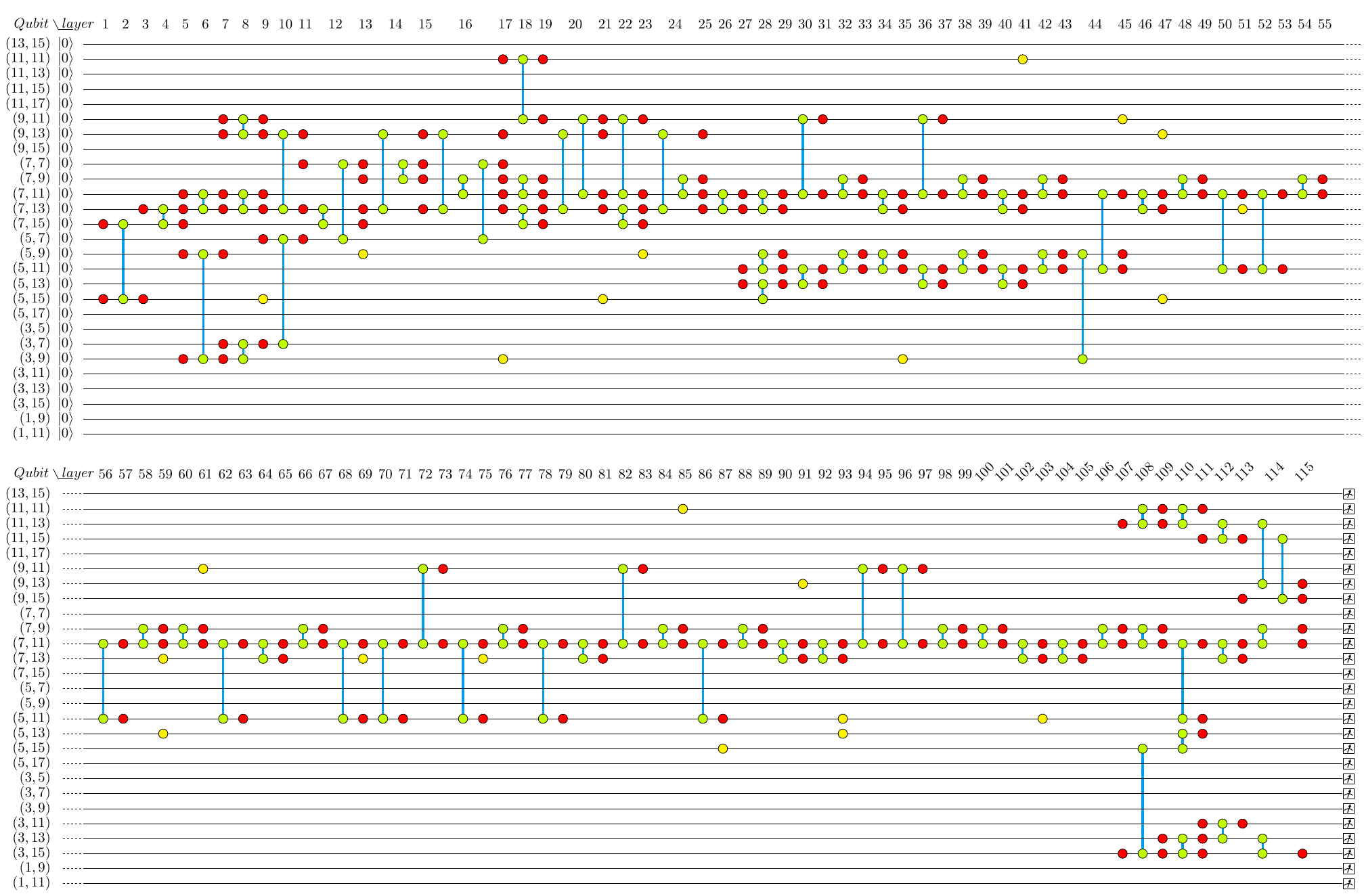}
\caption{\label{circuit_ground_state_2}\textbf{Experimental circuit for preparing the ground state of the Fibonacci string-net model}, which is obtained by variationally optimizing the F-move operations shown in the main text Fig.~2\textbf{b}. The whole circuit contains 115 layers, with each layer containing only single- or two-qubit gates that are applied simultaneously. 
}
\end{sidewaysfigure} 

\begin{figure}[htb]
\includegraphics[width=1\linewidth]{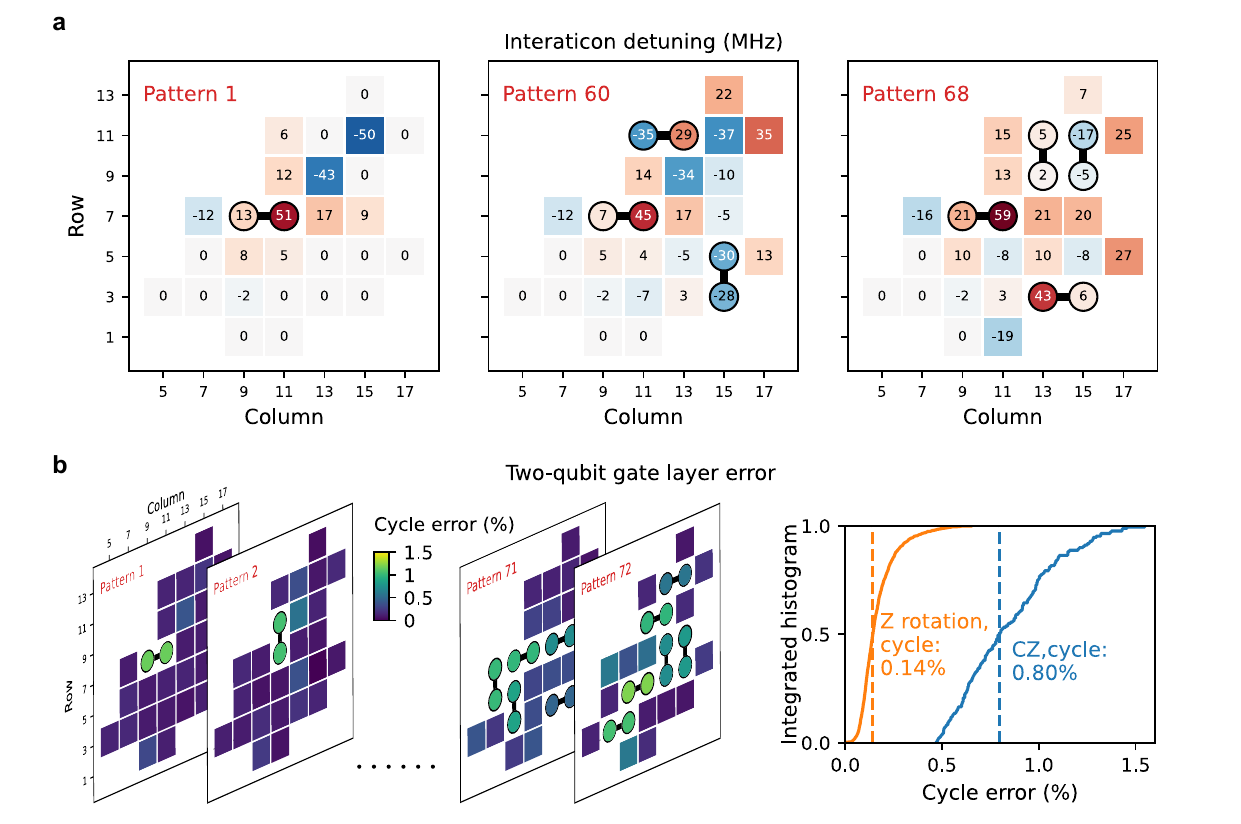}
\caption{\label{fig-property_tq}\textbf{Parameters of the two-qubit CZ gate layers.} \textbf{a}, Optimal interaction frequencies for three typical two-qubit gate layers with different patterns of parallel CZ gates applied on the highlighted qubit pairs (circles connected with solid lines). Displayed are the detunings of the qubits from their idle frequencies. We note that the parameters of a CZ gate may be different in different patterns. \textbf{b}, Cycle errors of the CZ gates (with each cycle consisting of a CZ gate and two single-qubit gates) and Z rotations (with each cycle consisting of two single-qubit gates). We characterize the single- and two-qubit gate errors with simultaneous XEB for all the 72 two-qubit gate layers used in the experiment, with each layer corresponding to a unique pattern of parallel CZ gates. The corresponding integrated histogram of the cycle errors for all the CZ gates and Z rotations are shown in the right. Dashed lines indicate the median values.}
\end{figure} 

We first optimize the single-qubit gate parameters, i.e., the idle frequencies, of all the qubits.
During the optimization procedure, we consider the coherence times and constrain the detunings between nearest and next-nearest qubits to construct the error model.
We assess the single-qubit gate performance by simultaneous cross-entropy benchmarking (XEB), and the results are shown in Fig.~\ref{fig-property_sq}j, with a median single-qubit gate Pauli error of 7e-4. In practice, we find that the optimized idle frequencies work for all the single-qubit gate layers in the experimental circuit, even though most of them only contain a specific part of parallel single-qubit gates.

The optimization of the two-qubit gate parameters is more subtle. Due to the engineered interactions and parasitic crosstalk, the optimal interaction frequencies of the CZ gates as well as the frequencies of the rest qubits in the system might be different for different two-qubit gate patterns. In this work, we optimize the gate parameters (including the idle frequencies of the rest qubits) for each different two-qubit gate layer in the experimental circuit, as shown in Fig.~\ref{fig-property_tq}a.
The resulting dynamical phases of the rest qubits are characterized with the method shown in Fig.~\ref{fig-calibrate_VZ} and canceled out with virtual Z gates.

The Pauli errors of all the CZ gates in the experimental circuit, characterized by the simultaneous XEB, are shown in Fig.~\ref{fig-property_tq}b, c, with a median value of 6.6e-3.

\begin{figure}[htb]
\includegraphics[width=1\linewidth]{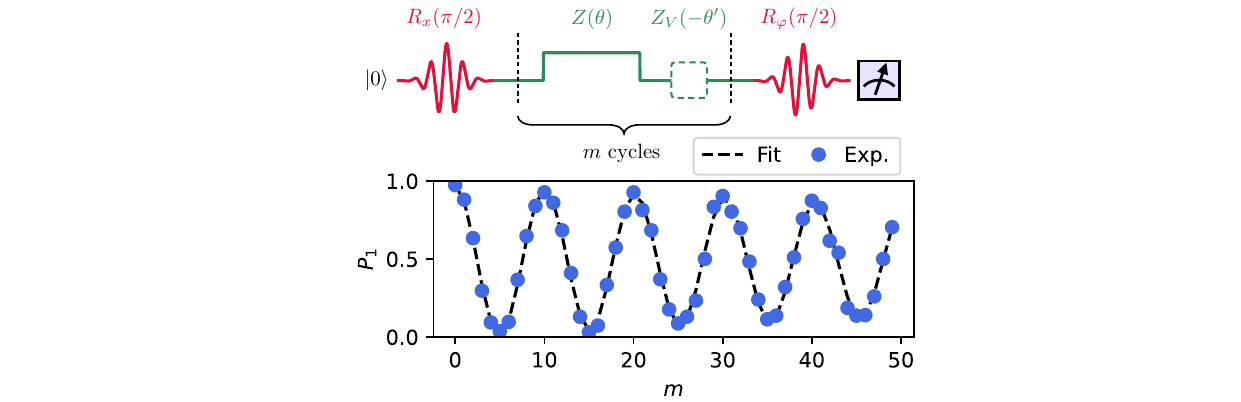}
\caption{\label{fig-calibrate_VZ}\textbf{Calibration of the compensated virtual Z gate in the two-qubit gate layer.} Top panel: The {pulse} sequence to amplify the tiny angle $\theta-\theta'$. We model the applied z pulse on the idle qubit during the execution of the two-qubit gate layer as a Z rotation $Z(\theta)$ which can be compensated with a virtual Z rotation $Z_V(-\theta')$ when $\theta-\theta'=0$. The final xy rotation $R_{\varphi}(\pi/2)$ is designed to detect the amplified angle $m(\theta-\theta')$ where the subscript $\varphi$ refers to an equatorial rotation axis that has an angle $\varphi$ with respect to the x-axis. In practice, $\varphi$ is chosen as $0.2m\pi$ where $m$ is the cycle number and the corresponding oscillation frequency of the measured probability of $|1\rangle$ is predicted as $f=0.1+(\theta-\theta')/(2\pi)$. Bottom panel: Example experimental data. We fit the data with the damped oscillation: $A\cdot \cos{(2 \pi f \cdot m+\Theta)} \cdot e^{-\Gamma m}+B$. Note that the calibration can be performed simultaneously on multiple qubits.}
\end{figure} 

\subsection{Topological entanglement entropy measurement}
For the string-net states studied in this experiment, the topological entanglement entropy can be calculated either from the von Neumann entropy
\begin{align}\label{eq:von-Neumann}
S_i = -\text{Tr}(\rho_i \ln \rho_i),
\end{align}
or the R\'enyi entropy of arbitrary order $\alpha$ (up to a negligible term of order $O(\exp(-L))$ with $L$ being the boundary length)
\begin{align}\label{eq:renyi}
S_{\alpha, i} = \frac{1}{1-\alpha}\ln\text{Tr}(\rho_i^\alpha),
\end{align}
where $\rho_i$ is the density matrix of the subsystem $i\in\{A, B, C, AB, AC, BC, ABC\}$~\cite{Flammia2009topo}. 
The difference of the entanglement entropy calculated by the two choices for the system considered in this experiment is shown in Fig.~\ref{fig-TEE}.
The experimental characterization of a density matrix is resource-consuming, with cost typically growing exponentially with the qubit number. Alternatively, we can measure the second-order R\'enyi entropy using the randomized measurement (RM) protocols~\cite{Tiff2019probing}  
without reconstructing the whole density matrix and thus reducing the required sampling number. Moreover, the RM method can extract all the entropies for different subsystems from the same data, which further reduces the required experimental resources. The core of RM is to obtain a set of classical representations of the target state by applying random unitary operations before measuring all the qubits on the computational basis. The second-order R\'enyi entropy can be predicted from the obtained classical data with high accuracy.

\begin{figure}[htb]
\includegraphics[width=1\linewidth]{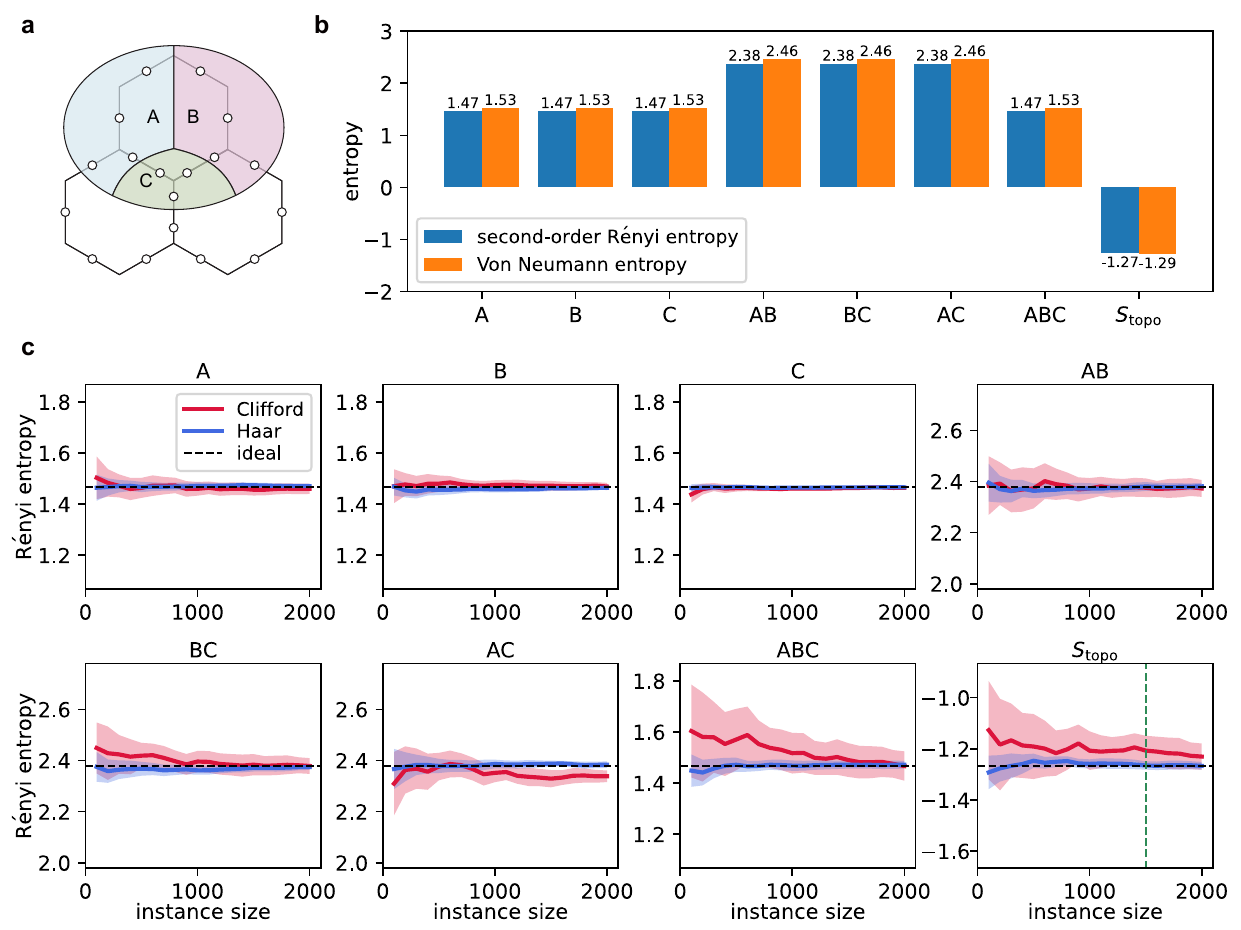}
\caption{\label{fig-TEE}\textbf{Noiseless numerical simulation of the topological entanglement entropy.
} \textbf{a}, The three simply connected subregions $A$, $B$, and $C$ that are used to extract the topological entropy. \textbf{b}, The comparison of the second-order R\'enyi entropies (blue) and Von Neumann entropies (orange) of all the subsystems and the corresponding topological entanglement entropy. The entropies of all the subsystems are calculated according to Eq.~\ref{eq:renyi}
 and Eq.~\ref{eq:von-Neumann} with the numerically simulated density matrices. \textbf{c}, The comparison of RM protocols using Haar-random rotations and random Clifford rotations. For each random unitaries instance size, we repeat Monte Carlo simulations 20 times to obtain the average value (solid line) and the standard deviation (light color region). 
The black horizontal dash line represents the ideal value of the second-order R\'enyi entropy calculated with Eq.~\ref{eq:renyi}. 
 The RM protocol using Haar-random rotations shows both less deviation from the ideal value and less statistical fluctuation. 
 We use 1500 Harr-random unitaries instances to obtain the second-order R\'enyi entropy in the experiment as indicated by the green vertical dash line in the right-down panel.
}
\end{figure} 

\begin{figure}[htb]
\includegraphics[width=1\linewidth]{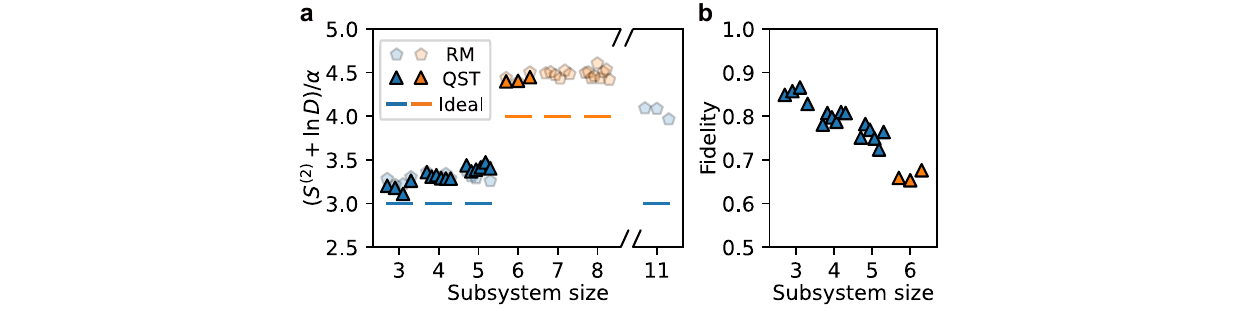}
\caption{\label{fig-TEE_state_tomo}\textbf{Experimental comparison of the second-order R\'enyi entropies extracted from RM and quantum state tomography (QST).} \textbf{a}, Distribution of the rescaled second-order R\'enyi entropies. We characterize the density matrices for all subsystems with qubit numbers less than seven using QST and then calculate the second-order R\'enyi entropies according to Eq.~\ref{eq:renyi}. The extracted data are displayed on top of the RM results as displayed in the main text Fig.~3\textbf{b}. As expected, the QST results are in good agreement with the RM results. \textbf{b}, The fidelities (defined as ${(\text{Tr}\sqrt{\sqrt{\rho_\text{ideal}}\rho_\text{exp}\sqrt{\rho_\text{ideal}}})}^2$) of the experimentally obtained density matrices $\rho_\text{exp}$ compared with the corresponding ideal ones $\rho_\text{ideal}$. 
}
\end{figure} 

The RM method requires analyzing the statistical correlations between the outcomes of measurements performed after random unitaries, and the estimator for the second-order R\'enyi entropy takes the following form:
\begin{align}\label{eq:supp-renyi_correlation}
\text{Tr}(\rho_i^2)=2^{N_i}\sum_{w,w'}(-2)^{-H(w, w')}\overline{P(w)P(w')},
\end{align}
where $N_i$ and $\rho_i$ are the qubit number and density matrix of the subsystem $i$. $w, w'$ are the binary strings and $H(w, w')$ is the hamming distance between them. $P(w)$ denotes the probability of observing $w$. The average is over different random unitaries in RM. 

In practice, there are two choices of random unitaries in RM, which can be tensors of either single-qubit Haar-random rotations or single-qubit random Clifford rotations, with the latter being equivalent to random Pauli basis measurement. 
The performance of RM depends on the quantum state under study as well as the selection of random single-qubit gates. 

For the ground state of string-net model in this work, we numerically study the efficiency of the two choices. As shown in Fig.~\ref{fig-TEE}c, the RM results with Haar-random rotations exhibit both less deviation from the ideal value and less statistical fluctuation than those with random Clifford rotations. In our experiment, we apply RM with 1500 instances of Harr-random unitaries to obtain the second-order R\'enyi entropy. 

We further verify the experimental results by performing quantum state tomography (QST) for the subsystem with qubit numbers less than 7, from which we obtain the density matrices and calculate the corresponding second-order R\'enyi entropies. In Fig.~\ref{fig-TEE_state_tomo}a, we plot the QST results on top of the RM results as displayed in the main text, which shows good agreement. The fidelities of the states are also shown in Fig.~\ref{fig-TEE_state_tomo}b.

\subsection{Readout error mitigation}\label{sec-readout_error_mitigation}

In this experiment, we use
the iterative Bayesian unfolding (IBU) method~\cite{DAGOSTINI1995487, Nachman2020unfolding} to mitigate the readout error. We first construct the response matrix $\mathcal{R}_{ij}=P(m=i|t=j)$ from the measured {single-qubit} readout fidelities~\cite{PhysRevLett.118.210504}, where $P(m=i|t=j)$ denotes the conditional probability of measuring the bitstring $i$ when the truth is the bitstring $j$ in the computational basis. Then we can extract the corrected probability by iteratively applying the following equation
\begin{align}\label{eq:IBU}
c_i^{n+1} = \sum_{j}\frac{\mathcal{R}_{ji}c_i^n}{\sum_{k}\mathcal{R}_{jk}c_k^n}m_j,
\end{align}
where $m_i$ and $c_i$ are the measured and corrected probabilities of bitstring $i$, respectively.
In practice, $c_i^0$ are initialized as a uniform distribution and the iteration number is 50.

Now we consider the error propagation in the IBU method. For a sampling experiment with $n$ repetitions, the random variable $X^i=(\sum_{k=1}^n{x_k^i})/n$ is an unbiased estimator for the measured probability $m_i$ where
\begin{align}\label{eq:sampled result}
x_k^i=
\begin{cases}
1, &\text{if sampled bitstring }i\\
0, &\text{otherwise}
\end{cases}.
\end{align}
But the intuitive estimator $X^iX^j$ for $m_im_j$ is actually biased since $E(X^iX^j)-Cov(X^i, X^j)=E(X^i)E(X^j)=m_im_j$. For convenience, we represent the map from the vector of measured probabilities $\Vec{m}$ to $c_i$ as $c_i=f_i(\Vec{m})$. Based on Taylor expansion, $f_i(\vec{X})$ can be approximated as $f_i(\vec{m})+\sum_k{J_{ik}}(X^k-m_k)$ where $\Vec{X}$ is the vector of $X^i$ and $J_{ik}=\frac{\partial f_i}{\partial x_k}\vert_{\Vec{x}=\Vec{m}}$ is the element of the Jacobian matrix. Then we can calculate the following expectation
\begin{align}\label{eq:error propagation}
E[f_i(\Vec{X})f_j(\Vec{X})]\approx&f_i(\Vec{m})f_j(\Vec{m})+f_i(\Vec{m})\sum_k{J_{ik}}E(X^k-m_k)+f_j(\Vec{m})\sum_l{J_{jl}}E(X^l-m_l)\notag\\&+\sum_{kl}{J_{ik}E[(X^k-m_k)(X^l-m_l)]J_{jl}}\notag\\
=&c_ic_j+\sum_{kl}{J_{ik}Cov(X^k, X^l)J^T_{lj}}.
\end{align}
Note that the covariance $Cov(X^k, X^l)$ can be estimated with the following unbiased estimator
\begin{align}\label{eq:covariance}
C_{kl}=\frac{\sum_i{(x_i^k-X^k)(x_i^l-X^l)}}{N(N-1)}
=\begin{cases}
[X^k-(X^k)^2]/(N-1),&k=l\\
-X^kX^l/(N-1),&k\neq l
\end{cases},
\end{align}
and the Jacobian matrix can be obtained with JAX~\cite{jax2018github}. From Eq.~\ref{eq:error propagation} and Eq.~\ref{eq:covariance} we can estimate $P(i)P(j)$ appeared in Eq.~\ref{eq:supp-renyi_correlation} with $f_i(\Vec{X})f_j(\Vec{X})-\sum_{kl}{J_{ik}C_{kl}J^T_{lj}}$ which converges faster than $f_i(\Vec{X})f_j(\Vec{X})$ and less repetitions are needed (see Fig.~\ref{fig-error_propagation}).

\begin{figure}[htb]
\includegraphics[width=1\linewidth]{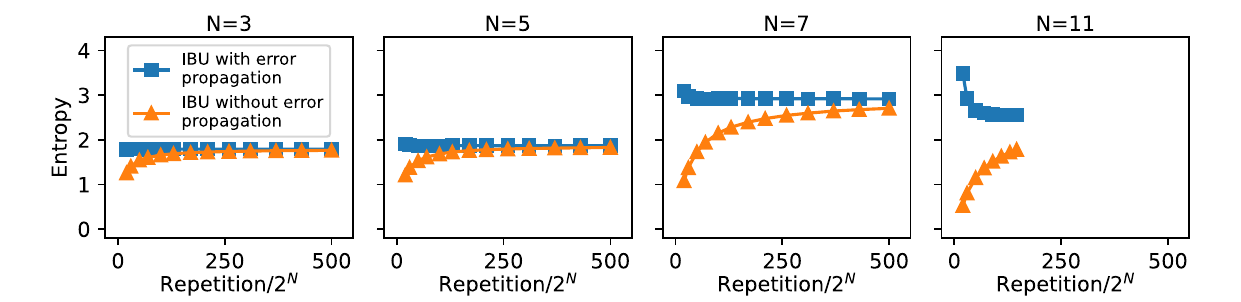}
\caption{\label{fig-error_propagation}\textbf{The second order R\'enyi entropy of different subsystem size N.} We compare the mitigated entropy using the IBU method with (blue square) and without (orange triangle) error propagation. Faster convergence is observed with error propagation and fewer repetitions are needed which are extremely resource-consuming. All data are extracted from the same experimental dataset with 300000 repetitions and {1500} instances.}
\end{figure}

\end{document}